\documentclass[acmsmall]{acmart}

\AtBeginDocument{%
  }

\setcopyright{acmlicensed}
\copyrightyear{2018}
\acmYear{2018}
\acmDOI{XXXXXXXX}

\usepackage{url} 
\usepackage{hyperref}
\usepackage{booktabs}
\usepackage{amsthm}
\usepackage{listings}
\usepackage[linesnumbered,ruled,vlined]{algorithm2e}
\usepackage{bm}
\usepackage{siunitx}

\usepackage[nounderscore]{syntax}
\usepackage{tcolorbox}
\usepackage[inkscapelatex=false]{svg}

\newcommand{\system}{\textsl{CPMMX}\xspace}
\newcommand{\sys}{\system}
\newcommand{\targetbug}{CPMM composability bug\xspace}
\newcommand{\targetbugs}{CPMM composability bugs\xspace}

\newcommand{\PP}[1]{\noindent\textbf{#1}.\xspace}

\newcommand{\answer}[1]{\vspace{5px}\noindent\fbox{
\parbox{\dimexpr\linewidth-5\fboxsep-2\fboxrule}{%
\noindent\textbf{#1}
}
}}
\newcommand{\exnum}{26\xspace}
\newcommand{\exprofit}{15.7K USD\xspace}

\usepackage{tikz}
\newcommand*\WC[1]{%
\begin{tikzpicture}[baseline=(C.base)]
\node[draw,circle,inner sep=0.2pt](C) {#1};
\end{tikzpicture}}

\usepackage{enumitem}
\newcommand{\squishlist}{
\begin{itemize}[noitemsep,nolistsep,leftmargin=*]
  \setlength{\itemsep}{-0pt}
}
\newcommand{\squishend}{
  \end{itemize}
}

\usepackage{multirow}
\newcommand{\cc}[1]{\mbox{\smaller[0.5]\texttt{#1}}}

\begin{document}

%%
%% The "title" command has an optional parameter,
%% allowing the author to define a "short title" to be used in page headers.
\title{Automated Attack Synthesis for Constant Product Market Makers}

\author{Sujin Han}
\affiliation{%
  \institution{KAIST}
  \city{Daejeon}
  \country{Republic of Korea}
}
\email{sujinhan@kaist.ac.kr}

\author{Jinseo Kim}
\affiliation{%
  \institution{KAIST}
  \city{Daejeon}
  \country{Republic of Korea}
}
\email{jinseo@kaist.ac.kr}

\author{Sung-Ju Lee}
\affiliation{%
 \institution{KAIST}
 \city{Daejeon}
 \country{Republic of Korea}
}
\email{profsj@kaist.ac.kr}

\author{Insu Yun}
\affiliation{%
  \institution{KAIST}
  \city{Daejeon}
  \country{Republic of Korea}
}
\email{insuyun@kaist.ac.kr}

\begin{abstract}
Decentralized Finance (DeFi) enables many novel applications that were impossible in traditional finances. However, it also introduces new types of vulnerabilities. An example of such vulnerabilities is a composability bug between token contracts and Decentralized Exchange (DEX) that follows the Constant Product Market Maker (CPMM) model. This type of bug, which we refer to as \targetbug, originates from issues in token contracts that make them incompatible with CPMMs, thereby endangering other tokens within the CPMM ecosystem. Since 2022, 23 exploits of such kind have resulted in a total loss of 2.2M USD. BlockSec, a smart contract auditing company, reported that 138 exploits of such kind occurred just in February 2023.

In this paper, we propose \system, a tool that automatically detects \targetbugs across entire blockchains.
To achieve such scalability, we first formalized \targetbugs and found that these bugs can be induced by breaking two safety invariants.
Based on this finding, we designed \system equipped with a two-step approach, called shallow-then-deep search.
In more detail, it first uses shallow search to find transactions that break the invariants.
Then, it uses deep search to refine these transactions, making them profitable for the attacker.
We evaluated \system against five baselines on two public datasets and one synthetic dataset.
% This evaluation demonstrated that \system outperforms all baselines in terms of recall, precision, and F1 score.
In our evaluation, \system detected 2.5x to 1.5x more vulnerabilities compared to baseline methods. It also analyzed contracts significantly faster, achieving higher F1 scores than the baselines.
Additionally, we applied \system to all contracts on the latest blocks of the Ethereum and Binance networks and discovered \exnum new exploits that can result in \exprofit profit in total.

\end{abstract}

\begin{CCSXML}
<ccs2012>
   <concept>
       <concept_id>10002978.10003022.10003023</concept_id>
       <concept_desc>Security and privacy~Software security engineering</concept_desc>
       <concept_significance>500</concept_significance>
       </concept>
   <concept>
       <concept_id>10011007</concept_id>
       <concept_desc>Software and its engineering</concept_desc>
       <concept_significance>500</concept_significance>
       </concept>
 </ccs2012>
\end{CCSXML}

\ccsdesc[500]{Security and privacy~Software security engineering}
\ccsdesc[500]{Software and its engineering}

\keywords{Smart Contract, Security, Composability, Exploit Generation}

\received{20 February 2007}
\received[revised]{12 March 2009}
\received[accepted]{5 June 2009}

%%
%% This command processes the author and affiliation and title
%% information and builds the first part of the formatted document.
\maketitle

% SUJ: Code figure setting

\definecolor{codegreen}{rgb}{0,0.6,0}
\definecolor{codegray}{rgb}{0.5,0.5,0.5}
\definecolor{codepurple}{rgb}{0.58,0,0.82}
\definecolor{backcolour}{rgb}{0.95,0.95,0.92}

\lstdefinestyle{mystyle}{
%    backgroundcolor=\color{backcolour},   
    commentstyle=\color{codegreen},
    keywordstyle=\color{magenta},
    numberstyle=\tiny\color{codegray},
    stringstyle=\color{codepurple},
    basicstyle=\ttfamily\tiny,
    % breakatwhitespace=true,         
    breaklines=false,
    % captionpos=b,                    
    % keepspaces=true                 
    numbers=left,                
    numbersep=2pt,                  
    showspaces=false,                
    showstringspaces=false,
    % linewidth=\textwidth,
    frame=single,              % Add a border around the code box
    framerule=0.05pt,             % Thickness of the frame
    framesep=7pt,              % Padding between the code and the frame
    aboveskip=0pt,             % No space above the code block
    belowskip=0pt,             % No space below the code block
    xleftmargin=0pt,           % No left margin
    xrightmargin=0pt,          % No right margin
    lineskip=0pt,             % Compact the lines
}

\lstset{style=mystyle}

% Define Solidity listing
\lstdefinelanguage{Solidity}{
  keywords={typeof, new, true, false, catch, function, return, if, in, while, else, for, 
 break, contract, address, private, public, internal, external, payable, modifier, uint, string, bool, int, bytes, mapping, event, emit, constructor, require, pragma, solidity, import, uint256},
  ndkeywords={class, export, boolean, throw, implements, import, this},
  ndkeywordstyle=\color{darkgray}\bfseries,
  identifierstyle=\color{black},
  sensitive=false,
  comment=[l]{//},
  morecomment=[s]{/*}{*/},
  % commentstyle=\color{purple}\ttfamily,
  stringstyle=\color{red}\ttfamily,
  morestring=[b]',
  morestring=[b]"
}

\section{Introduction}

Decentralized Finance (DeFi) provides new financial services using blockchain and smart contracts. 
These services use tokens, which are digital assets beyond native currencies on the blockchain. A key financial service smart contracts offer is Decentralized Exchanges~(DEX). Unlike Centralized Exchanges~(CEX), DEXes enable users to swap one asset for another without a central authority. Through DEXes, blockchain users can freely convert their assets, which provides fluidity in the blockchain economy. To enable swapping without an intermediary, most DEXes adopt the Constant Product Market Maker~(CPMM) model to automatically determine appropriate exchange rates. 

This DeFi ecosystem is often threatened by new types of vulnerabilities, one of which is \targetbugs. 
\targetbugs is a composability bug~\cite{clockworkfinance} arising from the interaction between a CPMM DEX and a token contract. 
This type of vulnerability allows an attacker to steal assets from a flawless DEX by exploiting a bug in a token contract. 
Recently, this type of vulnerability has been frequently exploited.
For instance, on January 20, 2023, an attacker leveraged BRA token's flawed tax mechanism to steal around 225K USD worth of digital assets from a DEX~\cite{bra}. Moreover, BlockSec, which is a renowned security auditing company, reported 138 attacks of such kind just in February 2023.

Several tools~\cite{echidna,ityfuzz,ye2024midas} have been developed for multi-contract bug detection, yet detecting \targetbugs remains challenging.
First, existing tools suffer from a large search space because they target multiple types of vulnerabilities.
Second, these tools are unsuitable for generating profitable transactions (i.e., end-to-end exploits) because most rely on coverage-guided fuzzing.
Coverage-guided fuzzing focuses on new coverage rather than making profits, which often requires multiple repetitions of internal calls.
Lastly, these tools mostly rely on timeouts without offering early termination. This property makes them unsuitable for scanning vulnerabilities across entire blockchains, where benign contracts are predominant. 

To address these challenges, we propose \sys, a tool that automatically detects \targetbugs and synthesizes profitable transactions for the detected vulnerabilities.
We first formalize \targetbugs and identify that these bugs are caused by two broken invariants between a token contract and a CPMM DEX.
Based on this, \sys employs a two-step approach, called \emph{shallow-then-deep search}, to detect \targetbugs.
In the shallow search, \sys attempts to discover transactions that break the invariants.
If such transactions are found, \sys refines them in the deep search to make them profitable.

We compared \sys against five baseline tools, Echidna~\cite{echidna}, Ityfuzz~\cite{ityfuzz}, DeFiTainter~\cite{defitainter}, Slither~\cite{slither}, and Mythril~\cite{mythril}, on two public and one synthetic datasets.
In our evaluation, \sys outperformed existing tools, detecting 2.5$\times$ and 1.5$\times$ more vulnerabilities on the two public exploit datasets. 
On the synthetic dataset, it achieved an F1 score of 0.97, compared to 0.66 for the next best tool, ItyFuzz~\cite{ityfuzz}.
Notably, \sys completed this analysis in 10 hours, which is 6.9$\times$ faster than ItyFuzz.
Furthermore, to demonstrate the effectiveness of \system in the real world, we ran it on Ethereum and Binance. It discovered \exnum profitable transactions, which can yield \exprofit profit.

To summarize, we make the following contributions: 
\begin{itemize}
    \item We formalize \targetbugs and identify two safety invariants that, when broken, allow an attacker to steal funds from DEXes.
    \item We design and implement \system that automatically detects \targetbugs across entire blockchains. It employs a novel approach, \emph{shallow-then-deep} search, to efficiently identify \targetbugs without false positives.
    \item We evaluate \system{} on several datasets and compare it with five baseline tools. Moreover, we demonstrate its effectiveness in the real world by running it on Ethereum and Binance; it identified \textbf{\exnum undiscovered vulnerabilities} that can yield \textbf{total \exprofit profit.}
\end{itemize}
\section{Background}

\PP{ERC20 tokens}
Tokens are digital assets on the blockchain. Among these tokens, the most commonly used are fungible tokens referred to as ERC20 tokens, defined through the Ethereum Request for Comments 20~(ERC20).\footnote{Although each blockchain may refer to them differently according to their protocol (e.g., BEP20 or TRC20), we collectively call them as ERC20 Tokens in this paper.} Native currencies (e.g., ETH in Ethereum or BNB in Binance) can also utilize ERC20 services through wrapper implementations, such as Wrapped Ethereum~(WETH) or Wrapped Binance Coin~(WBNB). 

The ERC20 standard requires a token smart contract to implement a set of Application Binary Interface~(ABI) consisting of 9 functions and 2 events. These functions are necessary for basic operations of tokens, such as \cc{transfer(address,value)} and \cc{balanceOf(address)}.
Such a uniform interface allows developers to build financial services, such as DEXes, for countless tokens without having to write custom code for each token. This design also increases the flexibility of ERC20 token implementation, as developers can freely implement each ABI function. However, it also increases the risk of potentially violating critical safety invariants within a service.
% , leading to security vulnerabilities.

% \input{figures/normal_cpmm}

\PP{Constant Product Market Maker model}
The Constant Product Market Maker~(CPMM) model is adopted by DEXes to automatically swap one ERC20 token for another ERC20 token at an appropriate exchange rate. The CPMM model states that, given a DEX holding $x$ amount of $X$ tokens and $y$ amount of $Y$ tokens, the product of $x$ and $y$ should remain the same (i.e., $x \times y = k$). When a user requests to swap $\Delta x$ amount of $X$ tokens for $Y$ tokens, the amount of $Y$ tokens the DEX returns, $\Delta y$, is calculated with the equation $(x + \Delta x) \times (y - \Delta y) = k$. Thus, any swap operation in a CPMM DEX can be represented as a movement along the curve $x \times y=k$.
 
 The majority of DEXes today charge a small percent fee for each exchange to provide profit for the liquidity providers who deposited the initial X and Y tokens.
 For example, the Uniswap protocol~\cite{uniswap_whitepaper}, which is the most widely used DEX, charges a 0.3\% fee for each exchange. As a result, most DEXes can be said to have adopted a modified version of the CPMM model, where the product of two assets slightly increases after each exchange (i.e., $x \times y \geq k$).
\section{Motivation}
\label{sec:motivation}

\vspace{-10px}
\begin{figure}[htbp]
        \begin{minipage}{0.48\textwidth}
        \centering
            \begin{lstlisting}[language=Solidity,breaklines=true]
uint public rewardRate = 5;
uint public percent = 10000;
uint public minAmount = 10000 * 1e18;
function giveReward(address receiver, uint amount) private {
  if (amount > minAmount) {
    rewardAmount = amount * rewardRate / percent;
    balances[receiver] += rewardAmount;
  }
}
function transfer(address sender, address receiver, uint amount) public {
  if (sender == DEX_ADDR) {
    giveReward(receiver, amount);
  } else if (receiver == DEX_ADDR) {
    giveReward(sender, amount);
  }
  balances[sender] -= amount;
  balances[receiver] += amount;
}
            \end{lstlisting}
            \vspace{-10px}
     \caption{Vulnerable code snippet in ANCH token.}
    \label{fig:anch-code}
        \end{minipage}
    \hspace{10px}
    \begin{minipage}{0.44\textwidth}
        \centering
% \centerline{\includesvg[width=0.8\textwidth]{figures/motivating_example.svg}}
\centerline{\includegraphics[width=0.8\textwidth]{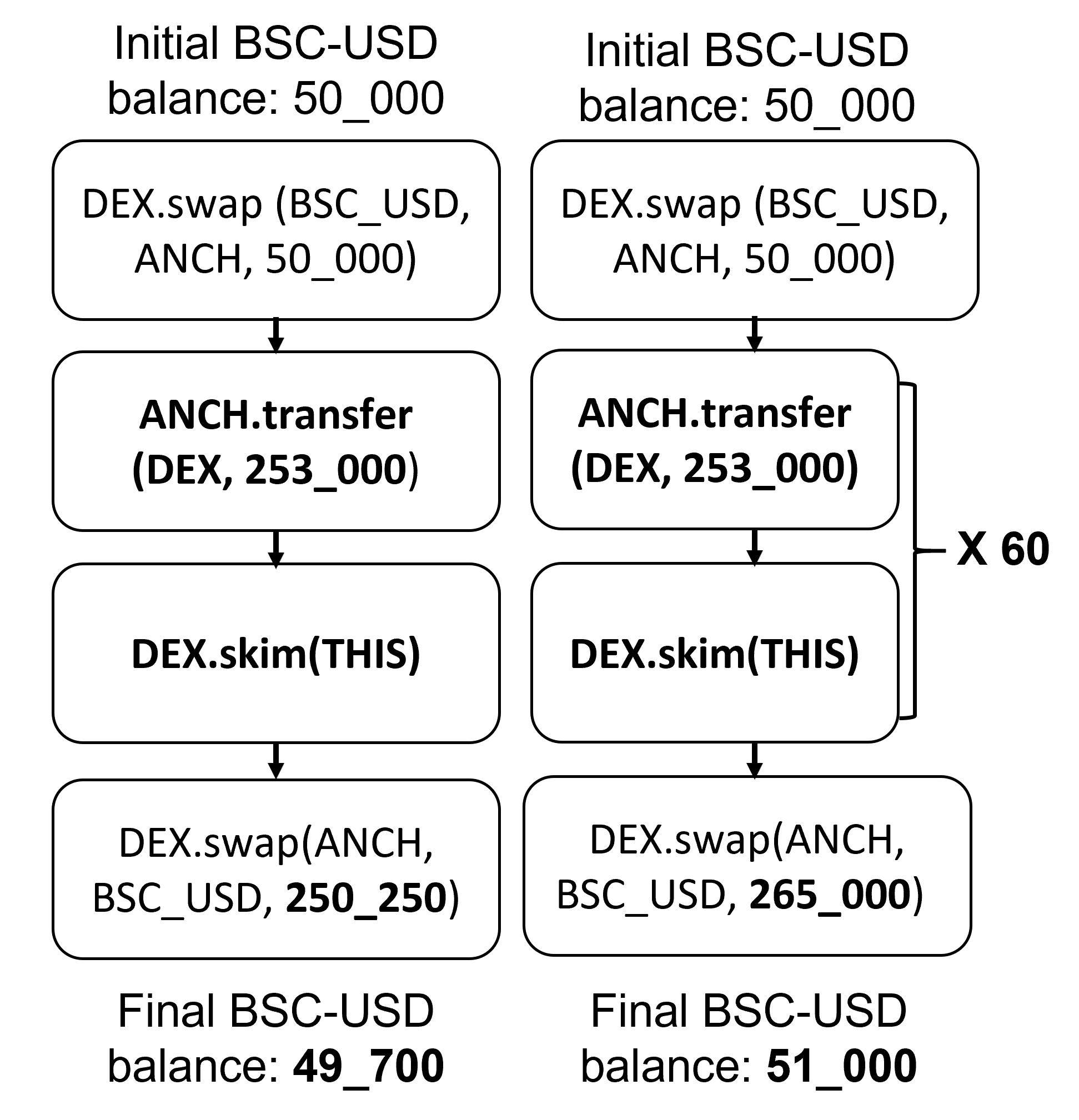}}
\caption{ANCH token exploit without repetition (left) and with repetition (right).}
\label{fig:motivating-example}
    \end{minipage}
    \vspace{-0.8cm}
\end{figure}

\subsection{Motivating Example}
\label{sec:motivation:example}
We present a motivating example that illustrates the challenges of detecting bugs in smart contracts and the need for a new approach to identify them effectively.
On August 9, 2022, an attacker exploited the ANCH token contract to extract approximately 19.9K USD worth of stablecoins from the ANCH-BSC-USD DEX~\cite{ancilia2022}. This happened because the attacker could increase their ANCH token balance without making any payment.
\autoref{fig:anch-code} shows a simplified version of the vulnerable code. In lines 11 and 13, the token contract checks whether the sender or receiver of the transfer is the DEX contract; if so, it rewards 0.05\% of the transfer amount to the receiver or sender, as shown in lines 6-7. Exploiting this behavior, the attacker accumulated many ANCH tokens to nearly drain the ANCH-BSC-USD DEX.

Unfortunately, the ANCH exploit cannot be easily detected with existing tools.
The key obstacle is that this bug needs to be triggered multiple times to generate profit. 
Existing fuzzers may produce test cases similar to the one shown on the left of \autoref{fig:motivating-example}, which triggers the bug but is not profitable. 
The attacker can gain rewards from \cc{ANCH.transfer(DEX, 253\_000)} and also \cc{DEX.skim(THIS)}.
Notably, \cc{skim} is a function that makes the DEX send extraneous tokens to the address given as the argument, internally calling \cc{ANCH.transfer(THIS, 253\_000)}.
Since the ANCH token rewards only 0.05\% of the transfer amount, receiving the reward twice is not enough to offset the swap fees, which are typically 0.3\%. Moreover, it is not desirable to flag any reward mechanism like this as vulnerable; many benign tokens exhibit similar behaviors to incentivize users.
Thus, building a profit-generating test case similar to the one on the right of \autoref{fig:motivating-example} is essential. However, existing tools struggle to generate such test cases because they require multiple repetitions of the reward-reaping call to generate profit. 
Existing tools like Echidna~\cite{echidna} or ItyFuzz~\cite{ityfuzz}, inspired by the success of coverage-guided fuzzing in traditional software (e.g., AFL~\cite{google_afl}), use guidance strategies that aim to maximize code coverage.
Unfortunately, such guidance strategies are ineffective at detecting this type of vulnerability because repetition does not improve code coverage but makes incremental changes in contract states.
On a more practical note, these tools require specific contracts for testing. Consequently, lesser-known contracts like ANCH could remain untested.

\subsection{Prevalence of CPMM Composability Bugs}
\label{sec:motivation:prevalence}

\vspace{-10px}
\begin{table}[htbp]
\caption{\targetbugs found in the DeFiHackLabs dataset.}
\vspace{-5px}
\label{tab:defihacklabs-explained}
\centering
\begin{minipage}{0.45\textwidth}
\resizebox{\textwidth}{!}{%
\begin{tabular}{@{}lcccc@{}}
\toprule
\multicolumn{1}{c}{\begin{tabular}[c]{@{}c@{}}Vulnerable\\ Token\end{tabular}} & \begin{tabular}[c]{@{}c@{}}Invariant \\ Broken\end{tabular} & \begin{tabular}[c]{@{}c@{}}Date \\ of Exploit\end{tabular} & \begin{tabular}[c]{@{}c@{}}Reported Loss\end{tabular} & \begin{tabular}[c]{@{}c@{}}Reported Loss\\ in USD\end{tabular} \\ \midrule
Wdoge & 1 & 2022/04/24 & 78.6 BNB & \SI{30.2}{K} \\
LPC & 2 & 2022/07/25 & \SI{45.1}{K} BSC-USD & \SI{45.1}{K} \\
ANCH & 2 & 2022/08/09 & \SI{19.9}{K} BSC-USD & \SI{19.9}{K} \\
XST & 2 & 2022/08/10 & 27.4 ETH & \SI{46.2}{K} \\
Shadowfi & 1 & 2022/09/02 & \SI{1.08}{K} BNB & \SI{300}{K} \\
PLTD & 1 & 2022/10/18 & \SI{24.5}{K} BSC-USD & \SI{24.5}{K} \\
HEALTH & 1 & 2022/10/20 & 16.6 BNB & \SI{4.54}{K} \\
AES & 1 & 2022/12/07 & \SI{61.6}{K} BSC-USD & \SI{61.6}{K} \\
BGLD & 1 & 2022/12/12 & 8.80 BNB & \SI{2.40}{K} \\
BRA & 2 & 2023/01/10 & 228K BSC-USD & \SI{228}{K} \\
Upswing & 1 & 2023/01/18 & 22.6 ETH & \SI{35.6}{K} \\
ThoreumFi & 2 & 2023/01/19 & \SI{2.26}{K} BNB & \SI{659}{K} \\
\bottomrule
\end{tabular}%
}
\end{minipage}%
\hspace{5px}
\begin{minipage}{0.48\textwidth}
\centering
\resizebox{\textwidth}{!}{%
\begin{tabular}{@{}lcccc@{}}
\toprule
\multicolumn{1}{c}{\begin{tabular}[c]{@{}c@{}}Vulnerable\\ Token\end{tabular}} & \begin{tabular}[c]{@{}c@{}}Invariant \\ Broken\end{tabular} & \begin{tabular}[c]{@{}c@{}}Date \\ of Exploit\end{tabular} & \begin{tabular}[c]{@{}c@{}}Reported Loss\end{tabular} & \begin{tabular}[c]{@{}c@{}}Reported Loss\\ in USD\end{tabular} \\ \midrule
SHEEP & 1 & 2023/02/10 & 9.54 BNB & \SI{2.93}{K} \\
Starlink & 1 & 2023/02/17 & 38.4 BNB & \SI{11.8}{K} \\
BIGFI & 1 & 2023/03/22 & \SI{30.3}{K} BSC-USD & \SI{30.3}{K} \\
GPT & 1 & 2023/05/25 & 155K BSC-USD & \SI{155}{K} \\
Bamboo & 1 & 2023/07/04 & 235 BNB & \SI{57.6}{K} \\
ApeDAO & 1 & 2023/07/18 & \SI{19.2}{K} BSC-USD & \SI{19.2}{K} \\
HCT & 1 & 2023/09/07 & 30.5 BNB & \SI{6.58}{K} \\
BFC & 1 & 2023/09/09 & \SI{42.3}{K} BSC-USD & \SI{42.3}{K} \\
pSeudoEth & 2 & 2023/10/08 & 1.44 ETH & \SI{2.34}{K} \\
TGBS & 1 & 2024/03/06 & 377 BNB & \SI{154}{K} \\
GHT & 1 & 2024/03/07 & 15.4 ETH & \SI{58.6}{K} \\
\bottomrule
\end{tabular}%
}
\end{minipage}
\end{table}
\vspace{-5px}

We refer to bugs that affect CPMM DEXes due to vulnerabilities in token contracts, like the ANCH exploit, as \emph{\targetbugs}.
Recently, attackers have frequently exploited \targetbugs to extract considerable amounts of tokens from DEXes. 
For example, BlockSec~\cite{blocksec}, a renowned smart contract auditing firm, reported 
that \targetbugs caused 138 exploits in February 2023.
We could also find \targetbugs in DeFiHackLabs~\cite{defihacklabs}, a public exploit replication dataset. The first and second authors manually inspected all exploits in the DeFiHackLabs dataset (as of March 2024) to create a subset where \targetbugs cause the exploits.
We found a total of 23 exploits and reported the details in \autoref{tab:defihacklabs-explained}. We categorized them based on the specific invariant broken, which is explained in \S\ref{sec:cpmm}. The cumulative financial loss of the 23 exploits is 2.2M USD.

\targetbugs pose a significant threat to financial assets stored on the blockchain because CPMMs are widely used to facilitate token exchanges and manage substantial financial assets. As of June 2023, DEXes following the CPMM model were reported to take up around 77\% of the market share, representing around 35.4 billion USD~\cite{coingecko}. Therefore, identifying \targetbugs is critical to ensuring the security of DeFi.

\subsubsection{Trend Analysis}

\begin{figure}[t!]
    \begin{minipage}{0.50\textwidth}
    % \centerline{\includesvg[width=\textwidth]{figures/proportion_cpmm_bugs_trend.svg}}
    \centerline{\includegraphics[width=\textwidth]{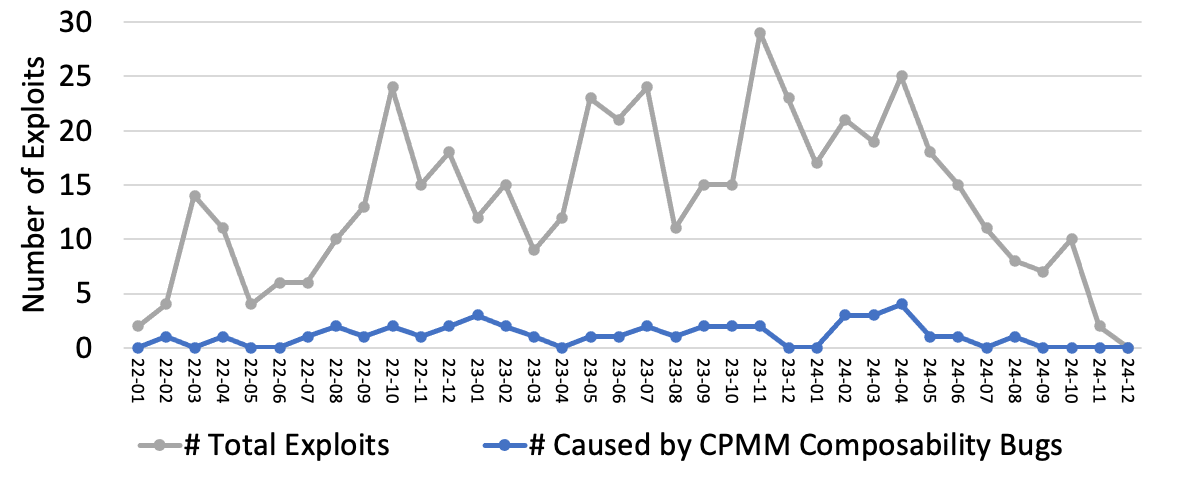}}
    \vspace{-0.4cm}
    \caption{Total exploits and \targetbug exploits per month in the DeFiHackLabs dataset.}
    \label{fig:proportion-trend}    
    \end{minipage}
    % \hfill
    \hspace{5px}
    \begin{minipage}{0.46\textwidth}
    % \centerline{\includesvg[width=\textwidth]{figures/financial_loss_cpmm_bugs_trend.svg}}
    \centerline{\includegraphics[width=\textwidth]{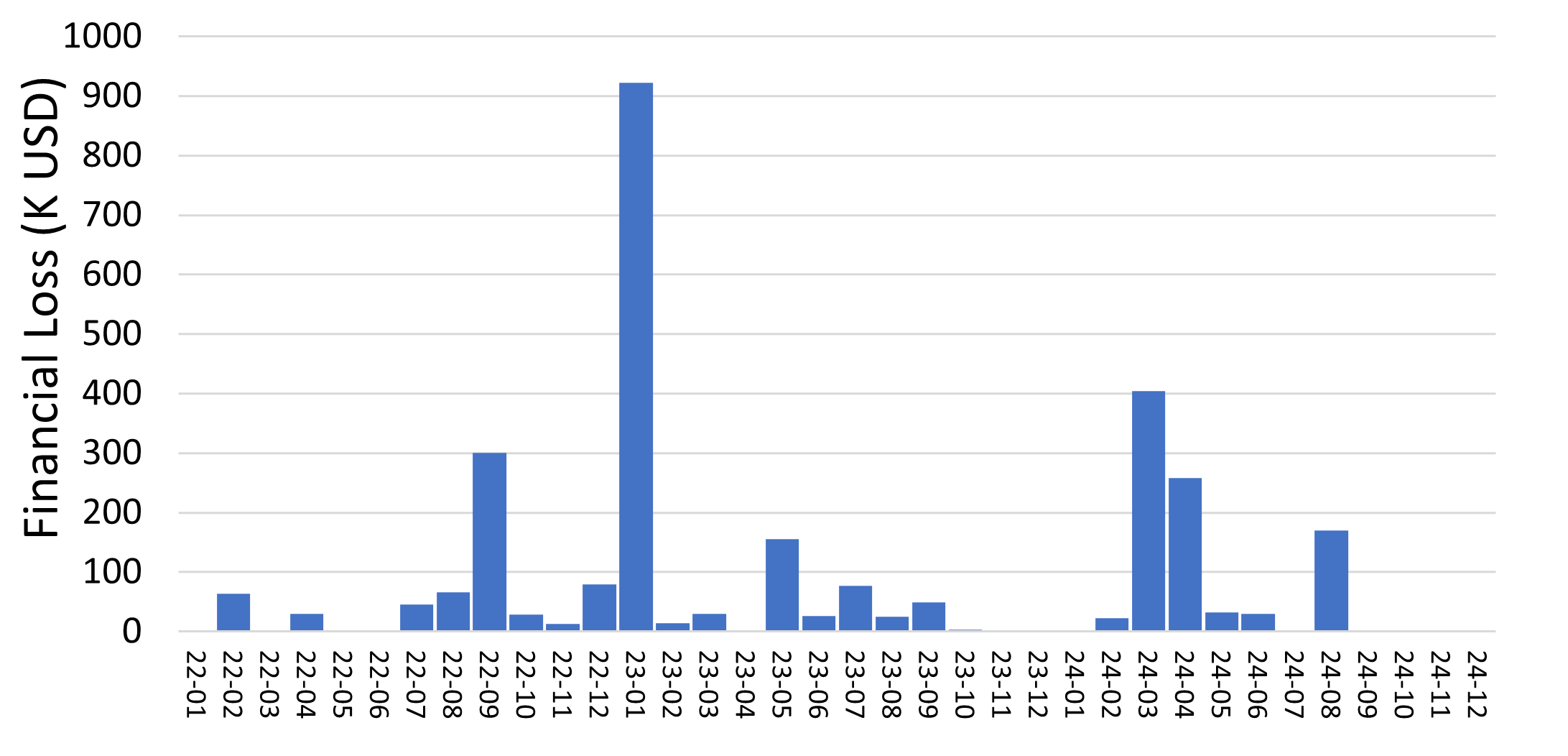}}
    \vspace{-0.4cm}
    \caption{Total financial loss per month due to \targetbugs in the DeFiHackLabs dataset.}
    \label{fig:fund-trend}
    \end{minipage}
    \vspace{-0.5cm}
\end{figure}

To illustrate the trend of \targetbugs, we report their monthly proportion and financial losses using the DeFiHackLabs dataset~\cite{defihacklabs}.
As shown in \autoref{fig:proportion-trend}, these bugs have been consistently reported from February 2022 to August 2024. 
We expect more \targetbugs to be reported for the last quarter of 2024, as incidents typically take a few months to be incorporated into the dataset.
As shown in \autoref{fig:fund-trend}, \targetbugs still incur significant financial losses.
For example, the ARK exploit (March 2024) caused a loss of 192K USD, while the IvestDao exploit (August 2024) resulted in 170K USD in losses.
\section{Goals and Approaches}

This section discusses the goals and approaches for building a tool, \sys, to detect \targetbugs across all blockchain contracts. 

\subsection{Detecting CPMM Composability Bugs Across Entire Blockchains}

As shown in \S\ref{sec:motivation:prevalence}, \targetbugs have been exploited repeatedly, posing significant threats to financial assets on the blockchain. However, existing tools are insufficient to detect \targetbugs at scale. First, as explained in \S\ref{sec:motivation:example}, the search strategies used by current tools are not well-suited to detect these bugs. Second, existing tools have limited scalability. Running them on the entire blockchain is computationally infeasible because they require significant computational resources to thoroughly test each contract. Therefore, there is a need for a tool capable of detecting \targetbugs across all smart contracts on the blockchain.

\PP{Our approach: formalizing \targetbugs and building a tool to detect them} To detect real-world vulnerabilities on a large scale, we decided to focus on \targetbugs and built a tool, named \sys, to automatically detect it.
\sys is designed to operate on the entire blockchain by minimizing computational costs. This is achieved by focusing the search space on areas likely to contain \targetbugs and implementing early termination for benign contracts. This was possible because \targetbugs exhibit properties that make them particularly suitable for automated detection.
In \S\ref{sec:cpmm}, we formalize this vulnerability and describe how we can detect it by checking invariant violations.

\subsection{Efficiently Detecting CPMM Composability Bugs}
Even after limiting our scope to \targetbugs, it is still challenging to analyze numerous smart contracts. 
As we can see from the motivating example in \S\ref{sec:motivation:example},
we need to build a complicated transaction with a long sequence of internal calls to exploit \targetbugs. 
Therefore, finding this bug by searching na\"ively requires an impractical amount of computation.

\PP{Our approach: shallow-then-deep search}
To address this problem, we propose a technique called \emph{shallow-then-deep search}. \sys searches for \targetbugs in two phases. 
The first phase, called shallow search, quickly identifies candidate contracts that may be vulnerable.
However, real-world smart contracts are diverse and complex, making distinguishing between signs of vulnerabilities and intended behaviors difficult.
To address this, \sys performs a second phase, called deep search, which explores the remaining contracts in more detail. In this phase, \sys generates a profitable transaction, which can be proof of the vulnerability. This approach allows us to analyze all smart contracts efficiently and detect \targetbugs.

\subsection{Minimizing False Positives}
 
Eliminating false positives is crucial for effectively detecting vulnerabilities across all smart contracts. Even with a low false positive rate, analyzing numerous smart contracts could still result in hundreds or thousands of false positive cases, placing a significant burden on analysts. 
 
\PP{Our approach: calculating profit using stablecoins and native currencies}
To address this, \sys automatically generates end-to-end profitable transactions. Unlike existing methods that calculate profit by approximating the value of coins (e.g., Midas~\cite{ye2024midas}), our approach adjusts transactions to ensure that all outcomes converge into tokens with relatively stable values (i.e., stablecoins or native currencies). \sys then computes the profit by evaluating the increase in these coins. This method is more reliable than previous approaches, which led to false positives due to noise in the value-based profit calculations.
In contrast, \sys allows users to analyze based on clear financial gains, minimizing ambiguity and false positives.
\section{Formalizing CPMM Composability Bugs}
\label{sec:cpmm}

In this section, we define and explain \targetbugs. First, we provide formal definitions in \S\ref{sec:cpmm:terminology}. Then, we explain how they can be utilized for profit with example exploits in \S\ref{sec:cpmm:type_one} and \S\ref{sec:cpmm:type_two}.

\subsection{Terminology}
\label{sec:cpmm:terminology}

\newcommand{\dex}{$DEX_{XY}$\xspace}

% Notation reference from Midas (ISSTA '24)

\PP{Notation} Given two ERC20 tokens, X token and Y token, a DEX following the CPMM model for the two tokens is denoted by \dex. We use the notation $BAL_X(S, entity)$ to denote the X token balance of $entity$ at blockchain state $S$. 
Furthermore, a transaction, $tx$, is a sequence of calls $c_1$$c_2$$c_3$ ... $c_n$ to contracts and are executed atomically in that order. 
Given initial blockchain state $S$, the blockchain state after executing transaction $tx$ is $S_{tx}$.

\PP{Profitability} We define profitability as gaining one token type in one transaction. 
Let $\mathbb{T}$ be the set of all tokens. 
A transaction $tx$ is profitable with respect to token $X$ if 

\begin{itemize}
\item $BAL_X(S, attacker) < BAL_X(S_{tx}, attacker)$ and 
\item $BAL_{Y}(S, attacker) \leq BAL_{Y}(S_{tx}, attacker)$ for all $Y \in \mathbb{T} \setminus \{X\}$.
\end{itemize}

\noindent We limit our scope to call sequences that can be executed in one transaction to exclude the impact of interest accumulation and other market players.
In a typical attack scenario, X token would be a coin with a relatively stable value, such as the wrapped native currency (e.g., WETH or WBNB) or a stablecoin (e.g., USDT). 
In addition, for a transaction to be truly profitable, the profit from the transaction should offset the gas fee involved in executing the transaction.
However, precise gas fee estimation is difficult because gas fees fluctuate based on several factors, including network congestion. 
Thus, \emph{for simplicity, we only consider transactions making profits over 1 USD as profitable}.

\PP{\targetbugs}
Consider a system composed of X token, Y token, and \dex following the CPMM model.
We \emph{define \targetbug as a bug in the Y token contract that enables an attacker to illegitimately extract X tokens from \dex to craft a profitable transaction with respect to the X token}.
Here, we assume that the X token contract and the \dex contract are free from vulnerabilities. This is a reasonable assumption given that X is a widely used stablecoin and \dex follows a standard implementation such as Uniswap. 

In general, users cannot gain X tokens by simply interacting with \dex; however, if the Y token contract contains a vulnerability or an incompatible behavior, an attacker can extract more than the initially inputted X tokens from \dex.
Under the assumption that the X token contract and the \dex contract are free from vulnerabilities, there are only two ways to extract more X tokens from \dex. That is, the attacker can either (1) decrease the Y token balance of \dex (i.e., $y$) or (2) increase $\Delta y$.
Recall that the formula for calculating the X token output (i.e., $\Delta x$) from a CPMM swap is $(x - \Delta x) \times (y + \Delta y) = k$. If $y$ decreases, $k (= x \times y)$ also decreases, increasing $\Delta x$. Similarly, when $\Delta y$ increases, $\Delta x$ can also be increased.
We further categorize \targetbugs into two types based on the invariant broken, as detailed in the following sections.

% Define 'invariant' environment
\newtheorem{invariant}{Invariant}

\subsection{Type 1: DEX Token Balance Decrease}
\label{sec:cpmm:type_one}

\vspace{-0.5cm}
\begin{figure}[htbp]
    \centering
    \begin{minipage}{0.62\textwidth}
        \centering
        % \includesvg[width=\textwidth]{figures/CPMM_pair_balance_decrease.svg}
        \includegraphics[width=\textwidth]{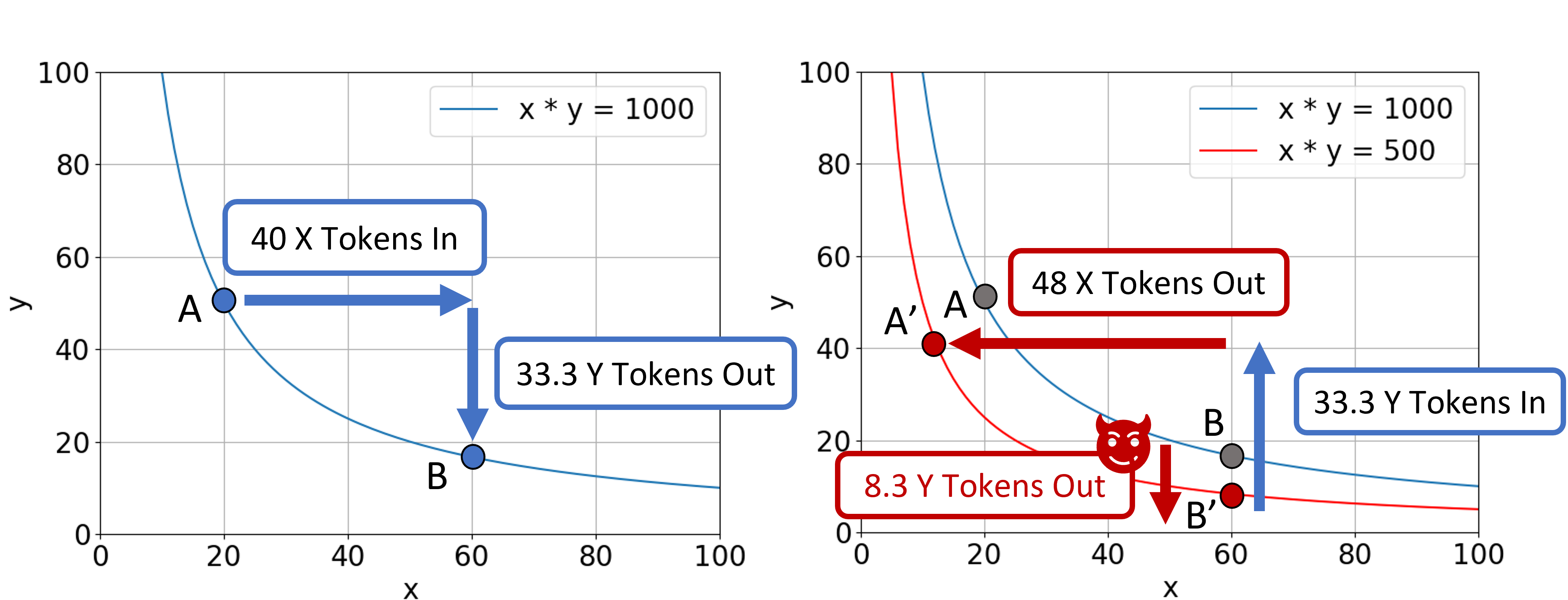}
        \vspace{-20px}
        \caption{Example attack scenario where the attacker is able to decrease Y token balance of $DEX_{XY}$.}
        \label{fig:pair-balance-decrease-CPMM}
    \end{minipage}
    % \hfill
    \hspace{5px}
    \begin{minipage}{0.33\textwidth}
        \centering
            \begin{lstlisting}[language=Solidity,breaklines=true]
function exploit() public {
  swapBNBtoShadowFi();
  // Decrease DEX ShadowFi balance
  ShadowFi.burn(DEX_ADDR, DEX_SHADOWFI_BALANCE-1);
  Pair(DEX_ADDR).sync();
  swapShadowFitoBNB();
}
            \end{lstlisting}
        \vspace{-5px}
        \caption{Simplified ShadowFi exploit.}
        \label{fig:shadowfi-exploit}
    \end{minipage}
    \vspace{-5px}
\end{figure}

The first type of \targetbug is a bug that allows an attacker to decrease the Y token balance of \dex without paying. An example scenario is shown in \autoref{fig:pair-balance-decrease-CPMM}. We assume that \dex has 20 X tokens and 50 Y tokens with $k = 1000$.
The attacker first swaps 40 X tokens for 33.3 Y tokens. 
% $(20 + 40) \times (50 - 33.3) \approx 1000$
Then, the attacker utilizes a \targetbug to decrease $y$ by 8.3 tokens, which decreases $k$ to 500; $60 \times (16.7 - \mathbf{8.3}) \approx \mathbf{500}$. This effectively shifts the swap curve inward. Since the attacker only decreased $y$ and $x$ remains the same, the next swap will happen at a point straight below the previous point on the updated swap curve (i.e., point $\text{B}^\prime$ instead of point $B$). The price of the Y token is greater at this point, allowing the attacker to swap the same amount of Y tokens for more X tokens (i.e., swapping to point $\text{A}^\prime$ instead of point A). The attacker ends up with 48 X tokens; $(60 - \mathbf{48}) \times (8.4 + 33.3) \approx 500 $, which is 8 X token profit. As illustrated with the example, when this invariant is broken, the attacker can extract X tokens from \dex.
% from $(x - \mathbf{\Delta x}) \times (y + \Delta y) = k$

\begin{invariant}[DEX token balance decrease]
Users should not be able to transfer or burn assets owned by a DEX without paying the DEX.
\end{invariant}

\PP{Real-world example} 
For instance, on September 2, 2022, an attacker exploited a vulnerability in the ShadowFi token to steal around 1078 BNB (worth around 301K USD)~\cite{shadowfi}. The vulnerability was that the ShadowFi token had a public \cc{burn} function. In DeFi, token burning refers to permanently removing tokens from circulation. In the exploit, the public \cc{burn} function allowed any user to remove ShadowFi tokens owned by any user. The simplified exploit is shown in \autoref{fig:shadowfi-exploit}. The attacker decreased the ShadowFi balance of the BNB-ShadowFi DEX using the \cc{burn} function (line 4) and updated the $k$ value of the DEX (line 5) to swap ShadowFi tokens for more BNB.

\subsection{Type 2: Attacker Token Balance Increase} 
\label{sec:cpmm:type_two}

% \vspace{-0.5cm}
\begin{figure}[htbp]
    \centering
    \begin{minipage}{0.62\textwidth}
        \centering
        % \includesvg[width=\textwidth]{figures/CPMM_attacker_balance_increase.svg}
        \includegraphics[width=\textwidth]{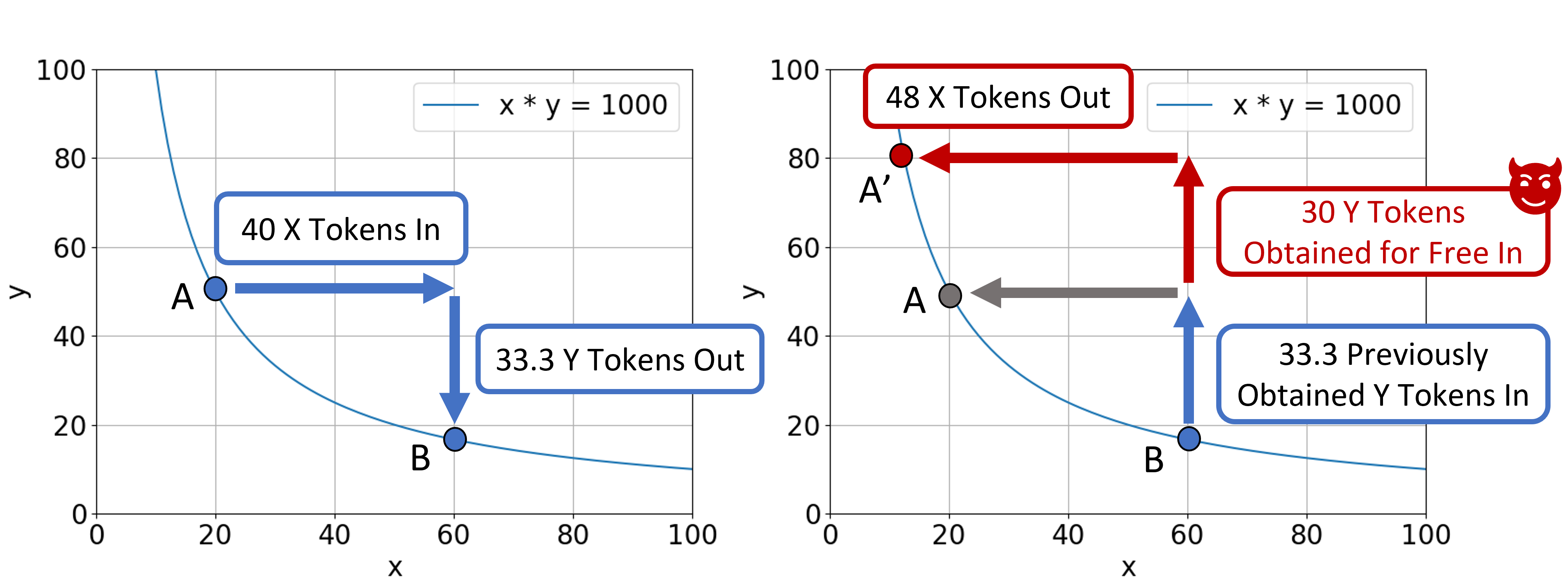}
        \vspace{-20px}
    \caption{Example attack scenario where the attacker is able to increase one's own balance of Y tokens.}
    \label{fig:attacker-balance-increase-CPMM}
    \end{minipage}
    \hspace{5px}
    \begin{minipage}{0.33\textwidth}
        \centering
            \begin{lstlisting}[language=Solidity,breaklines=true]
function exploit() public {
  swapBNBtoANCH();
  AMCH.transfer(DEX_ADDR);
  // Increase attacker ANCH balance
  for (uint i; i < 60; ++i) {
    Pair(DEX_ADDR).skim(DEX_ADDR);
  }
  Pair(DEX_ADDR).skim(THIS_ADDR);
  swapANCHtoBNB();
}
            \end{lstlisting}
    \vspace{-10px}
    \caption{Simplified ANCH exploit.}
    \label{fig:anch-exploit}
    \end{minipage}
    \vspace{-0.4cm}
\end{figure}

The second type of \targetbug is a bug that allows an attacker to gain Y tokens without cost.
An example scenario is shown in \autoref{fig:attacker-balance-increase-CPMM}. At point B, if the attacker can increase its own balance of token Y, then the attacker can gain more than the expected amount of X tokens (i.e., swapping to point A' instead of point A).
For the example, the attacker gains an additional 30 Y tokens, resulting in 8 X token profit; $(60 - \mathbf{48}) \times (16.7 + 33.3 + \mathbf{30}) \approx 1000$.
Hence, when \textbf{Invariant 2} is broken, the attacker can extract X tokens owned by \dex.

\begin{invariant}[Attacker token balance increase]
Users should not be able to obtain tokens traded in a DEX without cost.
\end{invariant}

\PP{Real-world example} 
For example, on August 9, 2022, an attacker leveraged ANCH token's reward mechanism to steal around 19.9K BSC-USD~\cite{ancilia2022}. The ANCH contract rewards users who buy or sell ANCH tokens from the ANCH-BSC-USD DEX. However, as shown in lines 4 to 7 in \autoref{fig:anch-exploit}, the attacker could illegitimately trigger the reward mechanism by abusing the \cc{skim} function in DEX that makes the DEX send extraneous tokens to any address given as the argument (\cc{skim} function exists to enable DEX to return leftover tokens from swaps to users). Utilizing this mechanism, the attacker could drain BSC-USD from the ANCH-BSC-USD DEX.
\section{Design}
\label{sec:design}

Based on the formalized model in \S\ref{sec:cpmm}, we design \system, an automatic tool that detects \targetbugs.
In this section, we describe its design in detail.

\subsection{Workflow}

\begin{figure}[htbp]
% \centerline{\includesvg[width=.85\textwidth]{figures/overview.svg}}
\centerline{\includegraphics[width=.85\textwidth]{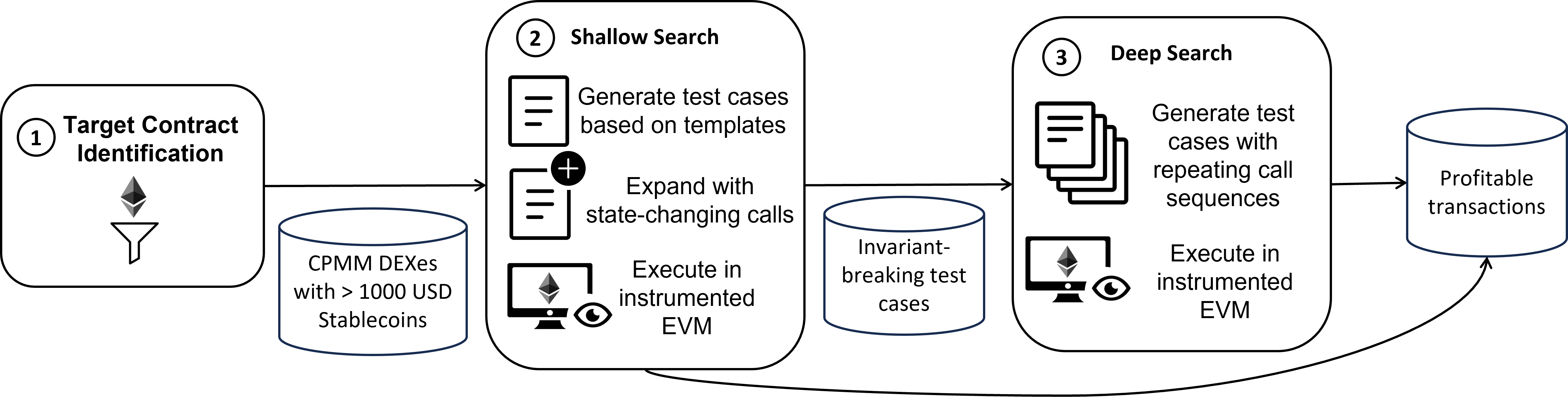}}
\vspace{-5px}
\caption{Overall workflow of \system{}.}
\label{fig:overview}
\vspace{-10px}
\end{figure}

 The overall workflow of \system is illustrated in \autoref{fig:overview}. 
 \WC{1} \system first identifies target contracts from the blockchain (\S\ref{sec:design:target-contracts}). 
 As \sys focuses on \targetbugs, it only considers contracts that follow the ERC-20 standard and are traded on a CPMM DEX. Moreover, it is not interesting to analyze contracts with insignificant financial value. 
 Therefore, \sys filters out DEXes with less than 1,000 USD worth of stablecoins or native currencies. 
 \WC{2} Then, \sys employs a two-phase search strategy, \textit{shallow-then-deep search}, to determine if the target contracts are vulnerable to \targetbugs. The shallow search phase uses predefined templates to find invariant-breaking transactions (\S\ref{sec:design:shallow-search}). If such transactions are already profitable, \sys outputs the transaction and flags the contract as vulnerable.
 On the other hand, if it cannot find any invariant-breaking transactions, \sys discards the contract (i.e., early termination).
 \WC{3} If \sys can only find invariant-breaking transactions, not profitable ones, it proceeds to the deep search phase (\S\ref{sec:design:deep-search}). In the deep search phase, the invariant-breaking call sequences are repeated to generate a profitable transaction for the target contracts. 

\vspace{-3px}
\subsection{Finding Target Contracts from the Blockchain}
\label{sec:design:target-contracts}

First, \sys scans the blockchain to find target contracts.
Our targets are token contracts traded in a CPMM DEX with a meaningful financial value.
To this end, \sys retrieves the addresses of all DEX contracts from the UniswapV2 factory contracts on both the Binance Smart Chain (BSC) and Ethereum networks. We used the most popular CPMM DEX platforms in each blockchain: PancakeSwap for BSC and UniswapV2 for Ethereum. 
Then, for each DEX contract, \sys collects the addresses of the tokens traded within the DEX and the balance of each token. From the list of all DEX contracts, \sys filters out contracts that do not meet the following criteria.

\begin{itemize}[noitemsep,nolistsep]
  \setlength{\itemsep}{-0pt}
    \item \PP{Exchangable to standard tokens} One of the tokens traded in the DEX is the wrapped native currency or a well-known stablecoin.
     \item \PP{Financially meaningful} The DEX contains over 1,000 USD worth of the wrapped native currency or a well-known stablecoin. 
     % We calculate the USD values by multiplying the token balance by the price of the token.
\end{itemize}

\noindent The DEX contracts that meet the above criteria and the associated tokens are then considered target contracts. 
As of October 2024, there were 1,674,491 PancakeSwap contracts and 371,380 UniswapV2 contracts. After filtering, we had 19,377 (1.16\%) and 28,800 (7.74\%) contracts, respectively. 

\vspace{-3px}
\subsection{Shallow Search for Finding Invariant-Breaking Transactions}
\label{sec:design:shallow-search}

\begin{table}[t!]
    \centering
    \caption{Templates for generating transactions by \sys whose bolded elements are repeatable.}
    \vspace{-10px}
    \resizebox{.8\columnwidth}{!}{%
\begin{tabular}{lll}
\toprule
    \textbf{Type}
        & \textbf{Testcase}
        & \textbf{Description} \\
    \midrule
%
    % \multirow{4}{*}{Cross-trading}
    Cross-trading
        & \cc{\textbf{transfer(this, transferAmount)}}
        & Send tokens to ourselves \\
        \cmidrule{2-3}
        & \begin{tabular}[c]{@{}l@{}} \cc{\textbf{transfer(DEX, transferAmount)}}\\ \cc{\textbf{DEX.skim(this)}} \\ \end{tabular}
        & Send tokens back and forth between DEX and ours \\
        \cmidrule{2-3}
        & \begin{tabular}[c]{@{}l@{}}\cc{transfer(DEX, transferAmount)}\\ \cc{\textbf{DEX.skim(DEX)}}\\ \cc{pair.skim(this)}\end{tabular} 
        & \begin{tabular}[c]{@{}l@{}}Send tokens to DEX, DEX sends tokens to itself, \\ DEX sends tokens to ours \end{tabular} \\
        \cmidrule{2-3}
        & \begin{tabular}[c]{@{}l@{}}\cc{transfer(DEX, transferAmount)}\\ \cc{\textbf{DEX.skim(DEX)}}\end{tabular}
        & Send tokens to DEX, DEX sends tokens to itself  \\
    \midrule
    Burn 
    & \cc{\textbf{burn(burnAmount)}}
    & Remove tokens from circulation \\   
    \cmidrule{2-3}
    & \cc{\textbf{burn(DEX, burnAmount)}} 
    & Remove tokens from DEX  \\
\bottomrule
\end{tabular}
}
\vspace{-3px}
    \label{fig:shallow_search}
    \vspace{-0.5cm}
\end{table}

\PP{Overview} After identifying target contracts, \sys begins the shallow search to find invariant-breaking transactions.
This works as follows.
First, \sys deploys and initializes the attacker contract with a stablecoin balance. Second, it generates a transaction consisting of calls $c_1$$c_2$$c_3$ ... $c_n$ where $c_1$ and $c_n$ are calls to swap the attacker contract's stablecoins to the target token via the victim DEX and vice versa, and calls $c_2$ ... $c_{n-1}$ are generated based on predefined templates.
Then, \sys executes the calls in an instrumented EVM environment and determines whether the transaction is profitable or breaks any invariants.
Finally, if no profit-generating transactions are found, \sys expands the test cases by incorporating state-changing calls.

\PP{Generating test cases with invariant-breaking templates}
% To build an interesting transaction,
To build transactions likely to break invariants,
\sys uses templates, which are described in \autoref{fig:shallow_search}, to generate transactions.
These templates have been inspired by patterns observed in previous exploits~\cite{blocksec,defihacklabs}.
From our analysis of these exploits, we observed that \targetbugs are typically caused by incentive mechanisms embedded in tokens. 
This is quite an intuitive observation because most tokens are designed to incentivize users to promote their tokens.
Some tokens incentivize trading by rewarding users on specific transfers or by employing deflationary mechanisms, removing tokens from circulation to increase their value.

To test incentive mechanisms of tokens, \sys uses two types of templates: cross-trading and token burning. First, \sys uses cross-trading templates to execute token transfers without causing a net change in balances.
As shown in \autoref{fig:shallow_search}, the first three strategies for cross-trading are trivial to see why they are cross-trading.
The fourth one is a bit more subtle, as the tokens are left in the DEX. However, since \sys immediately swaps all the tokens afterward ($c_n$), the remaining tokens in DEX are treated as inputs for the swap.
We also included the self-transfer template (i.e., the first row of \autoref{fig:shallow_search}). Although it does not directly interact with the victim DEX, it can still trigger invariant-breaking behaviors leading to \targetbugs and stablecoin losses.
Second, \sys also attempts to test token-burning functions. As mentioned, some tokens include explicit burn functions designed to remove tokens from circulation. To test this feature, \sys includes any function containing the term “burn” in its name.  
For argument values, we populate them with values commonly observed in exploits. The values are listed in \autoref{tab:arguments}.

Our approach is similar to other template-based searches, but unlike others, our templates have special, \emph{repeatable} elements. In \autoref{fig:shallow_search}, the repeatable elements are highlighted in bold. These elements involve the specific mechanisms that are likely to break invariants (e.g., transferring vulnerable tokens to/from DEX, burning tokens), and repeating them does not disrupt the cross-trading nature of the test cases. With these repeatable elements, we divide our search into two phases: shallow search and deep search. In the shallow search, \sys uses the templates without repeating them, enabling a quick assessment to discard contracts that do not break invariants.
% do not contain \targetbugs.
The deep search repeats the critical call sequences to potentially generate a profitable transaction. This design is crucial, as invariant-breaking sequences often need multiple executions to offset other costs and ultimately yield profit. However, incorporating repetition from the shallow search phase would unnecessarily increase the time needed to discard benign contracts.

\PP{Executing test cases in instrumented EVM}
After generating a test case, \sys runs it in an instrumented EVM to check whether the test case breaks any invariants from the formalized model in \S\ref{sec:cpmm}.
For that, \sys uses the instrumented EVM to track important state variables, such as the attacker's and DEX's token balances.
It also keeps a snapshot of the state variables before executing $c_2$. It compares them with the state variables after executing until $c_{n-1}$ to determine whether the call sequence $c_2$ ... $c_{n-1}$ breaks any invariants.
Finally, by comparing the amount of stablecoins before and after the transaction, \sys can determine whether the transaction is profitable.

\begin{figure}[t!]
    \begin{minipage}{0.55\textwidth}
    \centering
    \captionsetup{type=table}
    \caption{Arguments used in the shallow search phase.}
    \label{tab:arguments}
    \resizebox{\columnwidth}{!}{%
\begin{tabular}{ll}
\toprule
    \textbf{Parameter}
        & \textbf{Argument Values} \\
    \midrule
\cc{transferAmount} & 
\begin{tabular}[c]{@{}l@{}l@{}}
\cc{token.balanceOf(this)} \\ \cc{token.balanceOf(pair)} \\ \cc{0}
\end{tabular}  \\ \midrule
\cc {burnAmount} & \begin{tabular}[c]{@{}l@{}l@{}} \cc{token.balanceOf(pair) - 1}\\ \cc{token.totalSupply() - 2 * token.totalSupply() \textbackslash } \\ \quad \cc{/ token.balanceOf(pair)}~\cite{bigfi_exploit_2023}\end{tabular} \\
\bottomrule
\end{tabular}
}
    \end{minipage}
    \hfill
    \begin{minipage}{0.43\textwidth}
    \centering
    \input{figures/expand_with_state_changing_call}
    \captionsetup{type=figure}
    \vspace{-0.3cm}
    \caption{Incorporating a state changing call to a template testcase.}
    \label{fig:state-changing-calls}
    \end{minipage}
    \vspace{-0.5cm}
\end{figure}

\PP{Expanding test cases with state-changing calls}
If no profit-generating transactions are found,
\sys expands templates by incorporating state-changing calls. 
It considers two types of state-changing calls: small-amount transfers and no-argument function calls. 
Similar to the templates, this approach is inspired by token incentive mechanisms.
Some tokens include token price maintenance logic in transfer functions. Small amount transfers can trigger these functions without significantly impacting token balances. Thus, \sys includes small amount transfers (i.e., transfers with amount 0 or 1) in state-changing calls. On the other hand, some tokens have dedicated functions to trigger incentive mechanisms. While these functions may accept arguments, \sys only considers no-argument function calls. This design choice is based on two reasons. First, generating valid arguments for function calls is a nontrivial task that can greatly reduce efficiency. Second, restricting \sys to only no-argument function calls still detects most exploits. Hence, \sys includes only no-argument function calls in state-changing calls.
\autoref{fig:state-changing-calls} illustrates how a state-changing call is incorporated into a template test case.
Each template is combined with a state-changing call to create two additional test cases. In the first, the state-changing call is appended at the end of the repeating call sequence. In the second, it is inserted just before the last swap. The state-changing calls are deliberately inserted in these positions to preserve the self-trading nature of the test cases. These test cases are then executed and checked for broken invariants or profit generation. If no such transactions are detected, \sys terminates early.
Our approach is effective in practice; however, it is not exhaustive. This limitation is discussed in \S\ref{sec:discussion:limitations}.

\vspace{-3px}
\subsection{Deep Search for Generating Profitable Transactions}
\label{sec:design:deep-search}
As explained in \S\ref{sec:motivation:example}, an invariant-breaking transaction is insufficient to assure a contract is vulnerable.
This is because many benign tokens can exhibit similar behaviors but cannot be exploited to yield profit. 
Thus, to eliminate such false positives, \sys employs the deep search phase to craft a proof of vulnerability (i.e., a profitable transaction).

\PP{Executing test cases with repetitions}
If the shallow search phase identifies only invariant-breaking transactions, \sys attempts to generate a profitable transaction by exacerbating the invariant violation through repeating invariant-breaking call sequences. Thus, in the deep search phase, it generates test cases with increasing repetitions.
Among these, we need to allocate more resources to test cases that are more likely to yield profit.
To that end, \sys utilizes the final stablecoin balance after executing a test case as a guide to decide which test cases to prioritize.
If the final balance remains unchanged after increasing repetitions three times, it dismisses the test case, assuming further repetitions will not change the contract state.
If the final balance decreases, the test case is retained but with limited repetitions, as profitability is unlikely.
If the final balance increases, it continues increasing repetitions but still enforces a cap to avoid wasting resources on slow-growing profits.
In addition, it dismisses any test cases that result in transaction reverts.
\vspace{-3px}
\section{Implementation}

\sys was built on top of Foundry~\cite{foundry}, a well-known smart contract testing tool, and relies on it to set up the environment necessary for on-chain testing. 
Furthermore, \sys fetches contract ABIs from popular blockchain explorers: BSCScan~\cite{bscscan_website} and Etherscan~\cite{etherscan_website}.

\PP{State tracking and argument replacement} 
One challenge in testing smart contracts is generating test cases that do not revert. For instance, in a typical \cc{transfer(address, amount)} function, a transaction reverts if \cc{amount} exceeds the sender's balance. Therefore, \sys must generate a value less than the sender's balance. Moreover, if the token transfer involves an exclusive fee, the transfer amount must account for enough balance to cover the fee to avoid transaction reverting. Generating such valid arguments is challenging as token balances may change after each call execution. 
To address this issue, \sys leverages state tracking and runtime argument replacement. It monitors important state variables, such as token balances, and replaces arguments with the most current values during execution. Furthermore, when exclusive fees apply, it computes the appropriate transfer amount to ensure that the transaction proceeds without reverting. 
\vspace{-3px}
\section{Evaluation}
\label{sec:eval}

To evaluate \system{}, we answer the following research questions:
\begin{itemize}
    \item \textbf{RQ1:} How effective is \system at detecting \targetbugs?
    \item \textbf{RQ2:} How efficient is \system at detecting \targetbugs?
    \item \textbf{RQ3:} How significant are the techniques applied to \system?
    \item \textbf{RQ4:} How effective is \system at detecting undiscovered \targetbugs in the real world?
\end{itemize}

\subsection{Experimental Setup}

\subsubsection{Baseline Selection}
Among many existing tools, we selected five tools as baselines:
ItyFuzz~\cite{ityfuzz}, Echidna~\cite{echidna}, DeFiTainter~\cite{defitainter}, Slither~\cite{slither} and Mythril~\cite{mythril}.
We selected ItyFuzz, Echidna, and DeFiTainter as they support multi-contract analysis and can detect (a subset of) \targetbugs{}.
We also included Slither and Mythril, which do not support multi-contract analysis, to demonstrate that tools designed for single contracts are ineffective at detecting \targetbugs. 

In the following, we describe the configurations for each tool used in the evaluation.
Note that we tried our best to configure each tool for fair comparison.
Furthermore, DeFiTainter and Slither require source code analysis, so we could not run them for close-sourced contracts.
\begin{itemize}
\item \PP{ItyFuzz} We ran ItyFuzz with only the bug oracle that detects ERC20 token leaks. 
ItyFuzz also has a bug oracle for detecting token imbalances in DEXes (i.e., violations of \textbf{Invariant 1}), but these issues do not always lead to vulnerabilities. Therefore, we set up ItyFuzz to detect profitable transactions, similar to how \system operates as a whole.

\item \PP{Echidna} Echidna requires custom oracles to detect vulnerabilities.
Thus, we implemented an oracle that checks whether the attacker contract can get more native tokens after exchanging all ERC20 tokens for native ones.

\item \PP{DeFiTainter} DeFiTainter determines whether a given function contains a price manipulation vulnerability. 
Thus, we ran DeFiTainter for all public and external functions of a target contract and flagged it as vulnerable if any of the functions outputted a positive result. 

\item \PP{Mythril} Mythril has no detector for ERC20 token or ether leaks. However, other detectors might detect the programmatic error, leading to broken safety invariants for CPMMs. Thus, we manually validated results to check if Mythril can find the root cause of each exploit.

\item \PP{Slither} Since Slither includes many non-critical detectors, we ran Slither with only detectors that could be a potential root cause for \targetbugs (i.e. arbitrary-send-erc20, protected-vars, arbitrary-send-erc20-permit, arbitrary-send-eth, unchecked-transfer). Then, we manually validated its result to check if it could discover the root cause of each exploit. 
\end{itemize}

\subsubsection{Datasets}
We used three datasets for comparing \sys with existing tools: two public exploit datasets (DeFiHackLabs and BlockSec) and one custom-built dataset for evaluation (RealWorld-BSC). We use these datasets to answer 
\textbf{RQ1}, \textbf{RQ2}, and \textbf{RQ3}.
\begin{itemize}
\item \PP{DeFiHackLabs ($N=23$)} First, we use DeFiHackLabs~\cite{defihacklabs}, a public dataset for DeFi hacking incidents. This dataset is widely used for evaluating smart contract analysis tools~\cite{sun2024gptscan, zhang2023your, ye2024midas}. 
We used 23 exploits from this dataset that utilize \targetbugs as shown in \ref{sec:motivation:prevalence}.

\item \PP{BlockSec ($N=124$)} Second, we also used BlockSec~\cite{blocksec-twitter}, a public dataset containing 138 real-world exploits that involve breaking \textbf{Invariant 1}. Out of the 138 exploits, we use 124 exploits for this evaluation, as 14 are duplicate ones. 
Only one exploit, the SHEEP token exploit, is included in both the DeFiHackLabs and BlockSec datasets.

\item \PP{RealWorld-BSC ($N=244$)} Third, we constructed a dataset named RealWorld-BSC.
Unlike other datasets, we attempt to include both vulnerable and benign contracts to measure the real-world performance of each tool.
This dataset consists of 122 vulnerable contracts from BlockSec, which are deployed on the BSC, and the same number of benign contracts.
We use early-deployed DEX contracts with over 100 WBNB (worth around 60,000 USD) from PancakeSwap for benign contracts.
This is based on our assumption that tokens traded for a long time and holding substantial assets are less likely to be vulnerable.
\end{itemize}

In addition to these datasets, \system was also run on a large scale ($N>10,000$) on the Ethereum and Binance chains to evaluate its effectiveness in the real world (\textbf{RQ4}).

\subsubsection{Timeout and Number of Trials}
We used 20 minutes as the timeout for each contract, which is reasonably long enough if we consider the number of contracts to analyze in the real world. We ran fuzzing-based approaches (i.e., \system, ItyFuzz, and Echidna) three times for each case and reported the average values to avoid non-deterministic results from fuzzing.

\subsection{Effectiveness in Detecting Composability Bugs}
\label{sec:eval:effectiveness}

\begin{table}[htbp]
\caption{\targetbug{} detection rate of \system{} and baselines on the DeFiHackLabs dataset.}
\vspace{-10px}
\label{tab:defihacklabs-detection}
\centering
\resizebox{\columnwidth}{!}{
\begin{tabular}{@{}lccccccccccccccccccccccc|cc@{}}
\toprule
\multicolumn{1}{c}{} & \rotatebox{90}{AES} & \rotatebox{90}{ANCH} & \rotatebox{90}{ApeDAO} & \rotatebox{90}{Bamboo} & \rotatebox{90}{BFC} & \rotatebox{90}{BGLD} & \rotatebox{90}{BIGFI} & \rotatebox{90}{BRA} & \rotatebox{90}{GHT} & \rotatebox{90}{GPT} & \rotatebox{90}{HCT} & \rotatebox{90}{HEALTH} & \rotatebox{90}{LPC} & \rotatebox{90}{PLTD} & \rotatebox{90}{pSeudoEth} & \rotatebox{90}{Shadowfi} & \rotatebox{90}{SHEEP} & \rotatebox{90}{Starlink} & \rotatebox{90}{TGBS} & \rotatebox{90}{ThoreumFi} & \rotatebox{90}{Upswing} & \rotatebox{90}{Wdoge} & \rotatebox{90}{XST} & \rotatebox{90}{Total} & \rotatebox{90}{Recall} \\ \midrule
DeFiTainter & 0   & 0   & 1   & 0   & 0   & 0   & 0   & 0   & -   & -   & 0   & 0   & 0   & 0   & -   & 0   & 0   & 0   & 0   & -   & 0   & 0   & 0   & 1/19    & 0.05 \\
Echidna     & 0   & 0   & 0   & 0   & 0   & 0   & 0   & 0   & 0   & 0   & 0   & 0   & 0   & 0   & 0   & 0   & 0   & 0   & 0   & 0   & 0   & 0   & 0   & 0/23    & 0.00 \\
ItyFuzz     & 1   & 0   & 0   & 1   & 0.33& 0   & 0.33& 0   & 0   & 0   & 0.33& 0   & 1   & 0   & 1   & 0   & 0.33& 0   & 1   & 0   & 1   & 1   & 0   & 8.33/23 & 0.36 \\
Mythril     & 0   & 0   & 0   & 0   & 0   & 0   & 0   & 0   & 0   & 0   & 0   & 0   & 0   & 0   & 0   & 0   & 0   & 0   & 0   & 0   & 0   & 0   & 0   & 0/23    & 0.00 \\ 
Slither     & 0   & 0   & 0   & 0   & 0   & 0   & 0   & 0   & -   & -   & 0   & 0   & 0   & 0   & -   & 0   & 0   & 0   & 0   & -   & 0   & 0   & 0   & 0/19    & 0.00 \\ \midrule
Ours        & 1   & 1   & 0   & 1   & 1   & 1   & 1   & 1   & 0   & 1   & 1   & 1   & 1   & 1   & 1   & 1   & 1   & 1   & 1   & 1   & 1   & 1   & 1   & 21/23   & \textbf{0.91} \\ \bottomrule
\end{tabular}
}
\end{table}
\vspace{-15px}
\begin{table}[htbp]
\caption{\targetbug{} detection rate of  \system{} and baselines on the BlockSec dataset.}
\vspace{-10px}
\label{tab:blocksec-detection}
\resizebox{0.6\textwidth}{!}{
\begin{tabular}{@{}ccccccc@{}} 
 \toprule
 & DeFiTainter & Echidna & ItyFuzz & Mythril & Slither & Ours\\ \midrule
\textbf{Total} & 1/123 & 9/124 & 74/124 & 0/123 & 0/124 & \textbf{109/124} \\
\textbf{Recall} & 0.01 & 0.07 & 0.60  & 0.00 & 0.00 & \textbf{0.88} \\ \bottomrule
\end{tabular}
}
\vspace{-5px}
\end{table}

To compare the effectiveness of \system in detecting \targetbugs with existing tools, 
we measured the recall of \system and each baseline. 
%
%
% Describe results
\autoref{tab:defihacklabs-detection} and \autoref{tab:blocksec-detection} show the results of running \system and baselines on the DeFiHackLabs and BlockSec datasets, respectively. 
Note that we report the average detection rates for fuzzers, which may result in fractional values.

In summary, \system outperformed other tools in detecting \targetbugs{}.
In DeFiHackLabs dataset (\autoref{tab:defihacklabs-detection}), 
\system achieved the highest recall value of 0.91, while ItyFuzz had the second-highest recall value of 0.36.
DeFiTainter detected only one vulnerability out of 19 contracts, thus having a recall value of 0.05.
Other tools failed to detect any vulnerabilities.
In the BlockSec dataset (\autoref{tab:blocksec-detection}), \system also achieved the highest recall of 0.88, while ItyFuzz had the second-highest recall of 0.60.

\sys achieved significantly higher recalls than other tools by efficiently targeting areas of the search space likely to contain profitable exploits for \targetbugs.
Some exploits, such as ANCH, require repeated invariant-breaking call sequences to yield profit. These repetitions do not increase code coverage but gradually change contract states. Such scenarios are unlikely to be explored by coverage-guided fuzzers like Echidna and ItyFuzz, whereas \sys addresses this search space effectively through its deep search phase.
DeFiTainter was ineffective at detecting \targetbugs; it detects a different type of vulnerability --- price manipulation vulnerabilities --- and can only cover a subset of \targetbugs. Similarly, it is unsurprising that Mythril and Slither could not detect any \targetbugs. Mythril and Slither are static analyzers limited to analyzing individual contracts, whereas \targetbugs arise from the interaction between vulnerable tokens and DEX contracts.
The results indicate the need for a targeted approach to detect \targetbugs. 

\sys could not exploit ApeDAO and GHT because of their unique characteristics unlike other contracts.
In particular,  ApeDAO's fee mechanism is special as it is cheaper to pay fees in stablecoins rather than in ApeDAO tokens during DEX exchanges. 
As a result, we require a calculated stablecoin transfer to the DEX to exploit this behavior, which is not covered by \sys.
For GHT, attackers exploited the vulnerability in the same block where developers deposited GHT into the DEX.
Since \sys runs on finalized state variables and does not account for intra-block transactions, all its test cases reverted due to the absence of GHT tokens in the victim DEX. 
This limitation is further discussed in \S\ref{sec:discussion:limitations}. 

\answer{Answer to RQ1: \system outperforms existing tools in detecting \targetbugs.}

\subsection{Efficiency in Detecting CPMM Composability Bugs}

\subsubsection{Precision}

\begin{figure}[htbp]
    \begin{minipage}{0.43\textwidth}
    \centering
    \captionsetup{type=table}
    \caption{Performance metrics and running time comparison of \system and baselines on the RealWorld-BSC dataset at block 25543755.}
    \vspace{-10px}
    \label{tab:precision}
\resizebox{\textwidth}{!}{%
\begin{tabular}{@{}l|ccc@{}}
    \toprule
    & \textbf{Echidna} & \textbf{ItyFuzz} & \textbf{Ours} \\ \midrule

    \textbf{Precision} & 0.80 & \textbf{1.00} & \textbf{1.00} \\
    \textbf{Recall}    & 0.09 & 0.49 & \textbf{0.93} \\
    \textbf{F1 Score}  & 0.16 & 0.66 & \textbf{0.97} \\ \midrule

    \textbf{Vulnerable Time (min)}  & 2290 & 1707 & \textbf{150} \\
    \textbf{Benign Time (min)}      & 2425 & 2440 & \textbf{447} \\
    \textbf{Overall Time (min)}     & 4715 & 4147 & \textbf{597} \\ \midrule
    \textbf{Vulnerable Timeout \#}  & 111.33 & 62 & \textbf{6} \\
    \textbf{Benign Timeout \#}      & 119.33 & 122 & \textbf{16} \\
    \textbf{Overall Timeout \#}     & 230.66 & 184 &\textbf{22} \\ \bottomrule
\end{tabular}%
}
    \end{minipage}
    \hfill
    \begin{minipage}{0.55\textwidth}
    \centering
    \input{figures/blocksec_time}
    \vspace{-20px}
    \captionsetup{type=figure}
    \caption{Heatmap of time taken to detect \targetbugs in BlockSec dataset (seconds).}
    \label{fig:time-heatmap}
    \end{minipage}
\end{figure}

To evaluate the precision of \sys against other tools, we tested \sys, ItyFuzz and Echidna on the RealWorld-BSC dataset. Other tools were excluded because they could not detect any vulnerabilities in the BlockSec dataset. To simulate a scenario where these tools scan the entire blockchain for vulnerabilities and to ensure consistent contract states, we ran all experiments at block 25543755, which is one block before the earliest exploit in the BlockSec dataset.

\autoref{tab:precision} presents the results of running \system and baselines on the RealWorld-BSC dataset. 
For precision, \sys and ItyFuzz achieved the highest precision with 1.00. Echidna reported a few false positive cases. 
We reviewed these cases and found that Echidna incorrectly flagged reverting transactions as invariant violations. Specifically, when converting tokens back to native currencies for comparison, some transactions reverted, leading Echidna to incorrectly classify them as invariant violations. Since such transactions would not be executed in real-world scenarios, we classified them as false positives. 
Although ItyFuzz did not report false positives, it can identify much fewer \targetbugs compared to \sys.
Please note that the recall values differ from those in \autoref{tab:blocksec-detection}.
This is because in the previous evaluation, we used different block numbers for each contract to ensure that the contracts are exploitable;
however, in this evaluation, we used a single block number, 25543755.

\subsubsection{Time}
% Contextualize
To evaluate the efficiency of \sys in detecting \targetbugs, we compare its running time to that of ItyFuzz and Echidna. 
\autoref{tab:precision} shows the average running time of each tool on the RealWorld-BSC dataset. 
\sys was the most efficient, taking 597 minutes (around 10 hours) to test all 244 contracts. In comparison, Echidna and ItyFuzz took 4147 minutes (around 69 hours) and 4715 minutes (around 79 hours), respectively. 
The efficiency of \sys can be largely attributed to its ability to terminate early for contracts that do not exhibit invariant violations, resulting in only 16 out of 122 benign contracts reaching the timeout limit. 
In contrast, other tools run until timeout if they cannot detect any vulnerabilities. If the timeout had been extended, the running times of Echidna and ItyFuzz would have likely been even longer.
In addition, in \S\ref{sec:appendix:fee-on-transfer}, we evaluate \sys's performance with benign fee-on-transfer tokens, which are more challenging to differentiate from tokens with \targetbugs.

\begin{table}[htbp]
\caption{Average time taken by \system{} and ItyFuzz to detect bugs in the DeFiHackLabs dataset (seconds).}
\vspace{-10px}
\label{tab:defihacklabs-time}
\centering
\resizebox{\columnwidth}{!}{
\begin{tabular}{@{}lccccccccccccccccccccccc|c@{}}
\toprule
\multicolumn{1}{c}{} & \rotatebox{90}{AES} & \rotatebox{90}{ANCH} & \rotatebox{90}{ApeDAO} & \rotatebox{90}{Bamboo} & \rotatebox{90}{BFC} & \rotatebox{90}{BGLD} & \rotatebox{90}{BIGFI} & \rotatebox{90}{BRA} & \rotatebox{90}{GHT} & \rotatebox{90}{GPT} & \rotatebox{90}{HCT} & \rotatebox{90}{HEALTH} & \rotatebox{90}{LPC} & \rotatebox{90}{PLTD} & \rotatebox{90}{pSeudoEth} & \rotatebox{90}{Shadowfi} & \rotatebox{90}{SHEEP} & \rotatebox{90}{Starlink} & \rotatebox{90}{TGBS} & \rotatebox{90}{ThoreumFi} & \rotatebox{90}{Upswing} & \rotatebox{90}{Wdoge} & \rotatebox{90}{XST} & \rotatebox{90}{Average} \\ \midrule
ItyFuzz  & 508         & -           & -  & 17          & 86           & -           & 927         & -          & -  & -          & 177         & -          & 102          & -           & 26          & -           & 795         & -          & \textbf{15} & -           & 33          & \textbf{27} & -          & 246 \\ \midrule
Ours     & \textbf{41} & \textbf{38} & -  & \textbf{12}  & \textbf{61}  & \textbf{15} & \textbf{31} & \textbf{48} & -  & \textbf{19} & \textbf{30}  & \textbf{54} & \textbf{22}  & \textbf{65}  & \textbf{7}  & \textbf{299} & \textbf{11}  & \textbf{113} & \textbf{86} & \textbf{24}  & \textbf{10}  & 75 & \textbf{11} & \textbf{51} \\ \bottomrule
\end{tabular}
}
\end{table}

\sys also detects \targetbugs much faster than other tools.
\autoref{tab:defihacklabs-time} shows the average time taken to detect each vulnerability in the DeFiHackLabs dataset. The last column shows the average time taken to detect vulnerabilities. On average, \system took 51 seconds to detect vulnerabilities, around 4.82 times faster than the average time taken by ItyFuzz. 
Moreover, \autoref{fig:time-heatmap} is a visual representation of how fast each tool was at finding each vulnerability in the BlockSec dataset. Each cell in the heatmap contains the result for one vulnerability, thus a total of 124 cells per tool. The red color indicates that the tool took a long time (close to 1,200 seconds or 20 minutes) to detect the vulnerability, while the blue color indicates that the tool took a short time (close to 0 seconds) to detect the vulnerability. White cells indicate that the tool could not detect vulnerability for all three trials.
As shown in the figure, \system could detect most vulnerabilities quickly, while ItyFuzz detected vulnerabilities in varying time frames. 
On average, ItyFuzz and Echidna took 383 seconds and 178 seconds to detect vulnerabilities, respectively. Meanwhile, \system took only 16 seconds to detect vulnerabilities, around 24 times faster than the average time taken by ItyFuzz and 11 times faster than the average time taken by Echidna. 

This evaluation demonstrates that \sys effectively detects \targetbugs at scale. However, it does not establish that \sys is more scalable overall because ItyFuzz and Echidna are designed for comprehensive smart contract testing rather than for rapid identification of specific vulnerabilities (e.g., \targetbugs). Thus, they do not prioritize minimizing execution time.
The results instead suggest that a targeted approach, which efficiently identifies the existence of specific bugs, is more suitable for large-scale vulnerability detection. 

\answer{Answer to RQ2: \sys is the most efficient tool to detect \targetbugs. It requires the least running time and does not report any false positives.}

\subsection{Ablation Study}
To assess the design of \sys, we perform ablation studies to measure the impact of shallow-then-deep search (\S\ref{sec:eval:ablation:shallowdeep}) and runtime argument replacement (\S\ref{sec:eval:ablation:argument}).
Furthermore, in \S\ref{sec:appendix:no-dex-fee}, we present additional ablation studies to compare \sys with a commonly used auditing technique that modifies DEX contracts to remove fees.

\newcommand{\sysnorepeat}{\textsl{CPMMX-NoRepeat}\xspace}
\newcommand{\sysnoinvariant}{\textsl{CPMMX-NoInvariant}\xspace}

\subsubsection{Shallow-Then-Deep Search}
\label{sec:eval:ablation:shallowdeep}
To demonstrate the effectiveness of our approach, we conducted ablation studies to evaluate its impact on bug detection rate and measure its instruction coverage.

\PP{Bug detection}
We conducted an ablation study with two modified versions of \system: \sysnorepeat and \sysnoinvariant. \sysnorepeat does not utilize repetitions for test case generation (i.e., only shallow search). \sysnoinvariant generates test cases with a random number of repetitions and directly checks for profit generation (i.e., only deep search). 
Similar to previous evaluations, we ran \sysnorepeat and \sysnoinvariant three times. 

On average, \sysnorepeat detected 11 and \sysnoinvariant detected 16 out of 23 vulnerabilities in the DeFiHackLabs dataset, which is significantly lower than the 21 vulnerabilities detected by \system. Such an outcome is expected; \sysnorepeat cannot detect vulnerabilities that require repetition for profit. Meanwhile, \sysnoinvariant cannot efficiently allocate resources to function calls that are more likely to lead to exploits.

\vspace{-5px}
\begin{figure}[htbp]
% \centerline{\includesvg[width=\textwidth]{figures/code_coverage.svg}}
\centerline{\includegraphics[width=\textwidth]{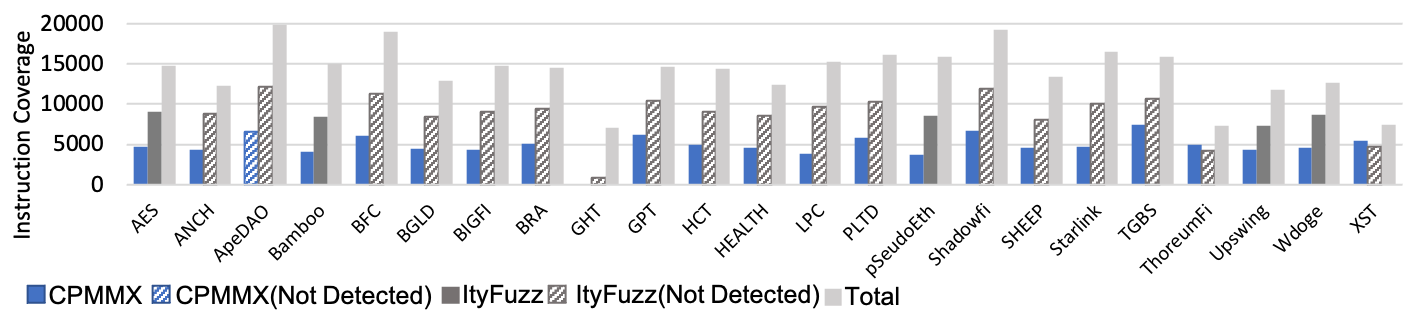}}
\vspace{-10px}
\caption{Code coverage reached by \sys and ItyFuzz on DeFiHackLabs dataset.}
\label{fig:code-coverage}
\vspace{-10px}
\end{figure}

\PP{Instruction coverage}
We also compare the instruction coverage of \sys with that of ItyFuzz. Total instruction coverage is measured as the sum of all instructions in the DEX contract and the corresponding token contracts. The results, presented in \autoref{fig:code-coverage}, reveal that while \sys detects more \targetbugs, ItyFuzz generally achieves higher instruction coverage. This suggests that ItyFuzz explores a broader range of functions than \sys, as it is designed to cover diverse vulnerabilities. In contrast, \sys is specially designed to detect \targetbugs. Its shallow search phase effectively filters out irrelevant functions, enabling a more targeted exploration of functions likely to reveal these vulnerabilities.

\newcommand{\sysnoargreplace}{\textsl{CPMMX-NoArgReplace}\xspace}

\subsubsection{Runtime Argument Replacement}
\label{sec:eval:ablation:argument}
To illustrate the necessity of \emph{runtime argument replacement}, we compare \sys to \sysnoargreplace, which does not employ this technique. 
More specifically, \sysnoargreplace stops state tracking after the initial swap to the target token to execute as if \sys generated test cases with fixed arguments.
In the DeFiHackLabs dataset, \sysnoargreplace detected only 5 out of 23 vulnerabilities. 
This is because it is highly prone to generating test cases that revert, as it cannot account for changing token balances. Furthermore, \sysnoargreplace cannot detect invariant violations that occur midway through a test case.

\answer{Answer to RQ3: The design of \sys, particularly shallow-then-deep search and runtime argument replacement, play a critical role in effectively detecting \targetbugs.}

\subsection{Effectiveness in the Real World}
\label{sec:eval:real-world}

\begin{table}[htbp]
\caption{Real-world exploits generated by \system. As these vulnerabilities have not been patched, we denote them with numbers to avoid providing details for exploitable vulnerabilities.}
\label{tab:generated-exploits-full}
\vspace{-10px}
\centering
\begin{minipage}{0.49\textwidth}
\resizebox{\textwidth}{!}{%
\centering
\begin{tabular}{@{}cccrr@{}}
\toprule
\begin{tabular}[c]{@{}c@{}}Exploit \\ Number\end{tabular} & \begin{tabular}[c]{@{}c@{}}Invariant \\ Broken\end{tabular} & Nework & \begin{tabular}[c]{@{}c@{}}Max Profit \\ in USD\end{tabular} & \multicolumn{1}{c}{\% Pair Asset} \\ \midrule
1 & 1 & BSC & 213.60 & 1.68 \\
2 & 2 & BSC & 398.00 & 0.74 \\
3 & 1 & BSC & 125.00 & 2.86 \\
4 & 1 & BSC & \textbf{4796.00} & \textbf{189.66} \\
5 & 1 & BSC & 282.76 & 2.40 \\
6 & 1 & BSC & 76.64 & 3.54 \\
7 & 1 & BSC & 2.14 & 0.05 \\
8 & 1 & BSC & 1.77 & 0.19 \\ 
9 & 1 & BSC & 1.80 & 0.25 \\ 
10 & 1 & BSC & 3.70 & 0.16 \\ 
11 & 1 & BSC & 1.43 & 0.0046 \\ 
12 & 2 & BSC & 614.00 & 1.14 \\ 
13 & 2 & ETH & 148.77 & 0.24 \\
\bottomrule
\end{tabular}
}
\end{minipage}
\hfill
\begin{minipage}{0.49\textwidth}
\resizebox{\textwidth}{!}{%
\centering
\begin{tabular}{@{}cccrr@{}}
\toprule
\begin{tabular}[c]{@{}c@{}}Exploit \\ Number\end{tabular} & \begin{tabular}[c]{@{}c@{}}Invariant \\ Broken\end{tabular} & Nework & \begin{tabular}[c]{@{}c@{}}Max Profit \\ in USD\end{tabular} & \multicolumn{1}{c}{\% Pair Asset} \\ \midrule
14 & 1 & ETH & 263.61 & 1.66 \\
15 & 2 & ETH & 23.49 & 0.28 \\
16 & 1 & ETH & 20.88 & 0.39 \\
17 & 1 & ETH & \textbf{4358.70} & \textbf{99.85} \\
18 & 1 & ETH & 13.05 & 1.90 \\
19 & 2 & ETH & 631.62 & \textbf{56.93} \\ 
20 & 1 & ETH & 46.98 & 0.71 \\ 
21 & 1 & ETH & 4.65 & 0.17 \\ 
22 & 2 & ETH & 5.90 & 0.44 \\ 
23 & 1 & ETH & 3.52 & 0.27 \\ 
24 & 1 & ETH & 21.82 & 1.81 \\ 
25 & 1 & ETH & 39.15 & 3.33 \\ 
26 & 1 & ETH & \textbf{3575.70} & \textbf{99.77} \\ 
\bottomrule
\end{tabular}
}
\end{minipage}
\vspace{-10px}
\end{table}

To demonstrate the effectiveness of \system in detecting undiscovered \targetbugs in the real world, we ran \system on the latest blocks of Ethereum and Binance. 
\autoref{tab:generated-exploits-full} contains the summary of profitable transactions generated by \system. \textit{Please note that we represent them with exploit numbers instead of token names or addresses because these vulnerabilities have not yet been patched.}
We discuss the issues with responsible disclosure for smart contracts in \S\ref{sec:discussion:responsible}.

In summary, \system could generate \exnum profitable transactions by exploiting \targetbugs{} in the real world, resulting in a total \exprofit profit.
To demonstrate the impact of each vulnerability, we report the maximum achievable profit in USD (column 4) and the ratio of this profit to the pair's balance before the exploit (column 5).
As \system halts when it finds a profit-generating transaction and does not proceed to maximize profit,
we manually adjusted some parameters of the exploit (e.g., initial token balance or the number of repetitions) to maximize the profit.
According to our experience, this profit maximization was straightforward for all exploits.

\system could generate four critical exploits that can drain more than half of the DEX's stablecoin balance (marked in bold in \autoref{tab:generated-exploits-full}).
Other attacks can also yield considerable profits (e.g., a few hundred dollars). We currently only consider single transaction exploits. However, if this vulnerability is exploited repeatedly in the long term, it could result in sustained profit and potentially drain all funds. We leave the development of such long-term exploit generation as future work.

\answer{Answer to RQ4: \system can generate impactful real-world exploits.}
\section{Case Study}

This section reports case studies for real-world \targetbugs that \sys detected.

\begin{figure}[htbp]
    \centering
    \begin{minipage}{0.37\textwidth}
        \centering
            \begin{lstlisting}[language=Solidity,breaklines=true]
function transfer ( address addr , uint amount ) external {
  if ( addr == DEX_ADDR ) {
    maintainPrice ();
  }
  // transfer tokens
  balances [ msg . sender ] -= amount ;
  balances [ addr ] += amount ;
}
function maintainPrice () internal {
  // decrease pair token balance by 10%
  balances [ DEX_ADDR ] =
  balances [ DEX_ADDR ] * 9 / 10;
}
            \end{lstlisting}
    \vspace{-10px}
    \caption{Vulnerable code snippet from \\Exploit 17.}
    \label{fig:case-one}
    \end{minipage}
    \hspace{15px}
    \begin{minipage}{0.54\textwidth}
        \centering
            \begin{lstlisting}[language=Solidity,breaklines=true]
function getRate () public {
  return totalTokenSupply / totalShareSupply ;
}
function transfer ( address addr , uint amount ) external {
  // transfer tokens
  uint shareAmount = amount / getRate ();
  shareBalances [ msg . sender ] -= shareAmount ;
  shareBalances [ addr ] += shareAmount ;
}
function maintainToken () external {
  // check that caller is a token owner
  require ( shareBalances [ msg . sender ] > minAmount );
  // proportionally decrease variables
  totalTokenSupply = totalTokenSupply * 9 / 10;
  totalShareSupply = totalShareSupply * 9 / 10;
  // award caller
  shareBalances [ msg . sender ] += awareAmount ;
}
            \end{lstlisting}
    \vspace{-10px}
    \caption{Vulnerable code snippet from Exploit 19.}
    \label{fig:case-two}
    \end{minipage}
    \vspace{-10px}
\end{figure}

% \vspace{-5px}
\subsection{Exploit 17: Breaking Invariant 1}

Exploit 17 is a real-world bug that violates \textbf{Invariant 1}.
This happens due to Token 17's deflationary mechanism.
\autoref{fig:case-one} shows the simplified version of vulnerable code from Token 17.
According to the CPMM model, whenever users sell Token 17 to the DEX, the price of Token 17 falls. 
To mitigate the price fall, Token 17 has a function that burns its share in the DEX whenever users sell Token 17 to the DEX (i.e., \cc{maintainPrice()} function in lines 9-13). 
Since this behavior can be triggered multiple times, an attacker can burn a significant portion of the DEX token balance, breaking \textbf{Invariant 1}.
The attacker can leverage this vulnerability to drain almost all stablecoins from the DEX, which is worth 4358.70 USD.
% , and leverage the vulnerability to drain stablecoins from the DEX. 
Interestingly, it is not profitable if we trigger this behavior only once, making it difficult for existing tools to detect this bug.
On the other hand, \sys can detect this bug by repeatedly triggering the behavior thanks to our two-step approach.

\vspace{-5px}
\subsection{Exploit 19: Breaking Invariant 2}

Exploit 19 is a real-world bug that breaks \textbf{Invariant 2}. %, which states that users should not be able to obtain tokens traded in a DEX without cost.
This token, henceforth Token 19, rewards users whenever they call a maintenance function.
% Unfortunately, this reward can be repeatedly reaped, allowing an attacker to accumulate a significant amount of Token 19.
\autoref{fig:case-two} shows the simplified version of Token 19.
This token manages its balances using two variables, \cc{totalTokenSupply} and \cc{totalShareSupply}.
As more users join the market for Token 19, the two variables will increase and may result in integer overflow. 
To prevent such a situation, Token 19 has to decrease the two variables periodically. 
Such maintenance function is implemented in lines 10 to 18 in \autoref{fig:case-two}.
Unlike other tokens that embed these functions in a commonly called function, such as \cc{transfer}, 
Token 19 adopts a different approach, incentivizing users to call this function directly.
However, the developers did not limit the number of times a user can call the function.
Thus, an attacker can repeatedly call this \cc{maintainToken()} to accumulate a significant amount of Token 19 without cost.
Note that, similar to our motivating example in \S\ref{sec:motivation:example}, the attacker has to reap the reward multiple times to offset costs, making it unlikely for existing tools to detect this bug.
We concluded that this vulnerability could be leveraged to drain around 57\% of relevant DEX's stablecoin balance, which is worth 631.62 USD.

\vspace{-5px}
\section{Discussion}
\label{sec:discussion}

\subsection{Responsible Disclosure for Smart Contract Vulnerabilities}
\label{sec:discussion:responsible}
We attempted to notify the token maintainers about bugs found by \sys but were unable to reach them.
We also reported our findings to CISA (Cybersecurity \& Infrastructure Security Agency), who recommended public disclosure of the bugs. However, we decided not to proceed with this recommendation, as anyone exploiting the vulnerabilities would directly cause financial harm to token holders.
In addition, we consulted with SEAL 911, a group of blockchain security researchers, but could not determine a safe and ethical approach for managing these vulnerabilities. As a result, the vulnerabilities remain unpatched. We hope that a safe and ethical approach will be established for addressing security issues in projects without active maintainers.

\vspace{-5px}
\subsection{Limitations and Future Works}
\label{sec:discussion:limitations}
\system has three limitations. First, it currently only supports Uniswap V2 DEXes. With further development, \sys can be extended to support other CPMM implementations, which we leave for future work. 
Second, \sys is constrained by the templates and state-changing calls used in test case generation. It fails to detect \targetbugs that do not conform to the templates, such as ApeDAO (\S\ref{sec:eval:effectiveness}). Furthermore, it does not leverage all available functions from token contracts for state-changing calls. It discards function calls with arguments to avoid the complexity coming from generating valid arguments.
To mitigate this limitation, we could extend templates and incorporate more state-changing function calls. Static analysis could help generate valid arguments. However, supporting a broader range of templates and functions would also increase the search space, potentially reducing efficiency.
Investigating the trade-off between efficiency and generalizability would be valuable in identifying the optimal balance.
Finally, \sys only supports \targetbugs, which can be represented by breaking two safety invariants.
In the future, it would be interesting to design a targeted approach like \sys for other vulnerabilities.

\vspace{-5px}
\subsection{Threats to Validity}
\PP{Internal threats} Since we suggest a new category of vulnerability, we could not evaluate our system on well-established datasets. Instead, we selected a subset from a popular dataset, DeFiHackLabs~\cite{defihacklabs}, as one of our evaluation datasets. To mitigate potential internal threats, the first and second authors independently selected the subset and discussed each exploit until reaching a consensus, minimizing selection bias and ensuring a consistent evaluation process.

\PP{External threats}
A potential external threat is the limited number of reported \targetbugs. To address this, we conducted an in-the-wild experiment as described in \S\ref{sec:eval:real-world}, which allowed us to validate \sys's performance against real-world contracts.

\section{Related Work}

Numerous tools detect smart contract vulnerabilities. Some utilize static analysis techniques, such as model checking~\cite{clockworkfinance} and symbolic execution~\cite{achecker, mossberg2019manticore}. While others utilize dynamic analysis techniques, most notably fuzzing~\cite{smartian, ityfuzz, echidna}. Recent works also utilize machine learning~\cite{chen2024improving, luo2024scvhunter}, including Large Language Models~\cite{sun2024gptscan, chen2024demystifying}.

\PP{Multi-contract vulnerability detection}
Several works were proposed to detect multi-contract vulnerabilities. Some focus on detecting commonly appearing ones, such as reentrancy and delegatecall-related vulnerabilities~\cite{xfuzz, liao2022smartdagger}, while some aim to detect a wide variety of vulnerabilities~\cite{ityfuzz, echidna}. In particular, ItyFuzz explores various combinations of contract states through fuzzing with snapshots, and Echidna uses a static analyzer, Slither, to extract useful information before fuzzing. Although \targetbugs can theoretically be detected with such methods, our evaluation indicates that a generic approach is ineffective. Since \targetbugs are closely tied to the business logic of contracts and sometimes require a long sequence of function calls for exploitation, a targeted approach is more suitable, as demonstrated by \system.

\PP{Automatic exploit generation} 
Some works propose systems that automatically generate exploits.
EthPloit~\cite{ethploit} generates exploits for single contracts based on fuzzing.
FlashSyn~\cite{flashsyn} utilizes counterexample-driven approximation to generate flashloan attacks. Gritti et al.~\cite{gritti2023confusum} designed a system that analyzes multiple contracts to automatically detect and exploit confused deputy vulnerabilities. \system pursues the same goal of exploit generation, but it targets a vulnerability that the aforementioned tools cannot detect. 
\section{Conclusion}
We proposed \system, a tool that can automatically detect and generate end-to-end exploits for \targetbugs.
For this, we first formalized \targetbugs and propose a two-step approach, called shallow-then-deep search, to detect them.
We evaluated \system against five baselines on three datasets and demonstrated that \system outperforms all baselines in terms of recall, precision, and F1 score.
Furthermore, we applied \system on the latest blocks of the Ethereum and Binance chains and discovered \exnum new exploits that can yield \exprofit total profit.
\section{Data Availability}
In support of the open science policy, we make the source code of \sys, datasets, and scripts to run experiments available at a public repository~\cite{kaist-hacking-CPMMX-2025}.

%%
%% The acknowledgments section is defined using the "acks" environment
%% (and NOT an unnumbered section). This ensures the proper
%% identification of the section in the article metadata, and the
%% consistent spelling of the heading.
\begin{acks}
This work was supported in part by the Korea-U.S. Joint Research Support Program funded by the Ministry of Science and ICT (MSIT) through the National Research Foundation of Korea (NRF) (RS-2022-NR119707) and the NRF grant funded by the Korea government (MSIT) (RS-2024-00337007).
\end{acks}

\appendix
\section{Appendix}

\newcommand{\sysnofee}{\textsl{CPMMX-NoFee}\xspace}
\newcommand{\sysnofeenorepeat}{\textsl{CPMMX-NoFeeNoRepeat}\xspace}

\vspace{-15px}
\begin{figure}[htbp]
    \begin{minipage}{0.44\textwidth}
    \centering
    \caption{Performance metrics and running time comparison of \sys and baselines on RealWorld-BSC-FOT dataset at block 25543755.}
    \label{tab:feeontransfer}
    \resizebox{\textwidth}{!}{%
    \begin{tabular}{@{}l|ccc@{}}
        \toprule
        & \textbf{Echidna} & \textbf{ItyFuzz} & \textbf{Ours} \\ \midrule
    
        \textbf{Precision} & 0.78 & 0.96 & \textbf{1.00} \\
        \textbf{Recall}    & 0.09 & 0.49 & \textbf{0.93} \\
        \textbf{F1 Score}  & 0.16 & 0.65 & \textbf{0.97} \\ \midrule
    
        \textbf{Vulnerable Time (min)}  & 2290 & 1707 & \textbf{150} \\
        \textbf{Benign Time (min)}      & 2411 & 2422 & \textbf{1867} \\
        \textbf{Overall Time (min)}     & 4701 & 4129 & \textbf{2017} \\ \midrule
        \textbf{Vulnerable Timeout \#}  & 111.33 & 62 & \textbf{6} \\
        \textbf{Benign Timeout \#}      & 119 & 119.67 & \textbf{83.67} \\
        \textbf{Overall Timeout \#}     & 230.33 & 181.67 &\textbf{89.67} \\ \bottomrule
    \end{tabular}%
    }
    \end{minipage}
    \hfill
    \begin{minipage}{0.55\textwidth}
    \centering
    \caption{Performance of \sys, \sysnofee, and \sysnofeenorepeat in detecting bugs in the DeFiHackLabs dataset.}
    \label{tab:nodexfee}
    \resizebox{\textwidth}{!}{%
\begin{tabular}{@{}l|cc@{}}
    \toprule
    & \textbf{\# Bugs Detected} & \textbf{Average Time (sec)}  \\ \midrule
    \sys & \textbf{21} & 51  \\
    \sysnofee & 20 & 43 \\ 
    \sysnofeenorepeat & 14 & \textbf{31} \\ \bottomrule
\end{tabular}%
}
    \end{minipage}
    \vspace{-20px}
\end{figure}

\subsection{Evaluation on Fee-on-Transfer Tokens}
\label{sec:appendix:fee-on-transfer}

To evaluate performance on challenging cases, we test \sys, ItyFuzz and Echidna on RealWorld-BSC-FOT, a modified version of the RealWorld-BSC in which all benign tokens are fee-on-transfer tokens. In the original RealWorld-BSC, only 22 out of 122 benign tokens have this property. Fee-on-transfer tokens deduct a portion of each transfer as a fee. 
In some cases, these fee mechanisms remove more tokens from DEXes than expected, causing \textbf{Invariant 1} violations. 
However, these tokens cannot be exploited to extract stablecoins from DEXes because they also implement safeguards.
This makes distinguishing between benign fee-on-transfer tokens and \targetbugs more difficult than differentiating non-fee tokens from \targetbugs.

As shown in \autoref{tab:feeontransfer}, although \sys continues to outperform the baselines, the gap in execution time has narrowed. \sys maintains a precision of 1.00, but its overall execution time increased from 597 seconds in \autoref{tab:precision} to 2017 seconds. This is because many benign fee-on-transfer tokens triggered \textbf{Invariant 1} violations.
Echidna and ItyFuzz demonstrated similar performances.
% to their results on RealWorld-BSC.

\vspace{-5px}
\subsection{Evaluation with DEXes Modified to Remove Fees}
\label{sec:appendix:no-dex-fee}

% \TODO{Include a comparative evaluation against simpler heuristic-based approaches, such as setting DEX fees to zero and performing swaps, to highlight the necessity of CPMMX's design}

An alternative approach to detect \targetbugs is executing template test cases with DEXes modified to remove fees.
% An alternative design for \sys to improve efficiency is to remove DEX fees.
This method is commonly used in real-world bug finding, as it can eliminate the need for repetitions required to identify vulnerabilities. For instance, in the case of the ANCH bug (\S\ref{sec:motivation:example}), \sys requires repetitions for the attacker to accumulate enough bonus to offset the DEX fee. If the DEX fee is removed, this bug can be detected without repetitions. To evaluate the effectiveness of this approach, we compare \sys with two variants using the DeFiHackLabs dataset: \sysnofee, which eliminates the DEX fee, and \sysnofeenorepeat, which removes both the DEX fee and the deep search phase for repetition.

As shown in \autoref{tab:nodexfee}, removing DEX fees resulted in lower bug detection rates. \sysnofee and \sysnofeenorepeat missed one more and seven more bugs compared to \sys.
\sysnofee was unable to detect the vulnerability in GPT. After patching the DEX bytecode to remove fees, all test cases for GPT reverted. This likely results from GPT internally calling DEX functions and expecting fee-based behavior.
\sysnofeenorepeat failed to detect \targetbugs in GPT as well as in six other tokens: BFC, BRA, HEALTH, PLTD, Starlink, and TGBS.
Even after removing DEX fees, these cases still required repetitions for profit. For instance, the PLTD token has both a fee and a bonus mechanism. Thus, the attacker must accumulate sufficient bonuses to offset PLTD's fees.
On average, \sysnofeenorepeat is the fastest at detecting bugs, followed by \sysnofee then \sys.
This suggests that removing DEX fees can be used as an optimization strategy to accelerate the bug detection process while maintaining reasonable accuracy.

\bibliographystyle{ACM-Reference-Format}
\bibliography{references}

\end{document}